\documentclass[12pt]{article}
\usepackage{epsfig,latexsym,amsbsy,amssymb,cite}

\begin{document}
\setlength{\unitlength}{0.2cm}

\title{
Dynamic Critical Behavior of an Extended Reptation Dynamics 
for Self-Avoiding Walks
}
\author{
  \\
  {\rm Sergio Caracciolo,$^{\rm a}$ }
  {\rm  Mauro Papinutto,$^{\rm b}$ and
          Andrea Pelissetto$^{\rm c}$}             \\[0.3cm]
  {\small\it ${}^{\rm a}$ 
      Dip. di Fisica dell'Universit\`a di Milano, I-20133 Milano,}
  \\[-0.2cm]
  {\small\it 
      Sez. INFN di Pisa, and NEST-INFM, I-56100 Pisa,
      ITALY} 
        \\[-0cm]
  {\small\it ${}^{\rm b}$ 
      Dip. di Fisica dell'Universit\`a di Pisa and Sez. INFN di Pisa,
      I-56100 Pisa, ITALY} 
        \\[-0cm]
  {\small\it ${}^{\rm c}$ 
      Dip. di Fisica dell'Universit\`a di Roma ``La Sapienza"}\\[-0.2cm]
  {\small\it  and
      Sez. INFN di Roma I, I-00185 Roma, ITALY} 
        \\[0.2cm]
{\small E-mail: {\tt Sergio.Caracciolo@sns.it}, }
   \\[-0.2cm]
{\small
                  {\tt papinutt@cibs.sns.it},  
                  {\tt Andrea.Pelissetto@roma1.infn.it} }\\
}
\vspace{0.5cm}

\maketitle
\thispagestyle{empty}   

\vspace{0.2cm}

\begin{abstract}
We consider lattice self-avoiding walks and 
discuss the dynamic critical behavior of two dynamics that use local and 
bilocal moves and generalize the usual
reptation dynamics. We determine the integrated and exponential 
autocorrelation times for several observables, perform a dynamic 
finite-size scaling study of the autocorrelation functions, and 
compute the associated dynamic critical exponents $z$.
For the variables that describe 
the size of the walks, in the absence of interactions we find
$z \approx 2.2$ in two dimensions and $z\approx 2.1$ in three dimensions.
At the $\theta$-point in two dimensions we have $z\approx 2.3$.
\end{abstract}

\clearpage

\newcommand{\be}{\begin{equation}}
\newcommand{\ee}{\end{equation}}
\newcommand{\beq}{\begin{equation}}
\newcommand{\eeq}{\end{equation}}
\newcommand{\bea}{\begin{eqnarray}}
\newcommand{\eea}{\end{eqnarray}}
\newcommand{\<}{\langle}
\renewcommand{\>}{\rangle}
\newcommand{\bi}{\begin{itemize}}
\newcommand{\ei}{\end{itemize}}
\newcommand{\bc}{\begin{center}}
\newcommand{\ec}{\end{center}}

\def\spose#1{\hbox to 0pt{#1\hss}}
\def\ltapprox{\mathrel{\spose{\lower 3pt\hbox{$\mathchar"218$}}
 \raise 2.0pt\hbox{$\mathchar"13C$}}}
\def\gtapprox{\mathrel{\spose{\lower 3pt\hbox{$\mathchar"218$}}
 \raise 2.0pt\hbox{$\mathchar"13E$}}}

\newcommand{\R}{\hbox{{\rm I}\kern-.2em\hbox{\rm R}}}
\newcommand{\reff}[1]{(\ref{#1})}
\def\smfrac#1#2{{\textstyle\frac{#1}{#2}}}

\section{Introduction} \label{sec1}

The lattice self-avoiding walk (SAW) is a well-known model for the 
critical behavior of a homopolymer in a 
solvent \cite{deGennes_book,desCloizeaux-Jannink_book} 
and it has also been extensively used in the study of several
properties of heteropolymers \cite{Lau-Dill_89,Sali-etal_94}. 
The earliest simulations either used a local dynamics
\cite{Verdier-Stockmayer_62} in which, at each step, a small part of the walk
(usually 2-4 consecutive beads) was modified, or the so-called reptation
dynamics \cite{Kron_65,Kron-etal_67,Wall-Mandel_75,Mandel_79}.
All these algorithms are however nonergodic
\cite{Madras-Sokal_87,Kron-etal_67,Wall-Mandel_75,Madras-Slade_book,
foot1} 
and only a limited 
fraction of the phase space is visited. Note that, contrary to some claims
in the literature, the deviations are sizeable even for very short walks 
if one is interested in low-temperature properties, i.e. 
polymers in the compact phase
or heteropolymers near the folding temperature
\cite{Karplus-Shakhnovich_book,Bryngelson-etal_95,Grassberger_book}. 
For instance---see footnote 4 in Ref. \cite{Caracciolo-etal_2000}---if 
one uses the Verdier-Stockmayer algorithm
\cite{Verdier-Stockmayer_62}
in two dimensions, one does not sample 3.2\%, 1.4\%, 5\% of the most compact 
configurations for $N=11,13,15$.
These ergodicity problems 
can be solved by using a different ensemble 
\cite{BF_81,AC_83,ACF_83,BS_85}, chain-growth algorithms 
\cite{Suzuki_68,Redner-Reynolds_81,Grassberger_97,GN-conf},
or nonlocal algorithms
\cite{Lal_69,MacDonald-etal_85,Madras-Sokal_88,Madras-etal_90,%
Caracciolo-etal_92,Kennedy_01}.
However, in the presence of interactions, nonlocal algorithms become 
inefficient since nonlocal moves generate new walks with large energy 
differences and thus they will be rejected making the dynamics very slow. 
Chain-growth algorithms may work much better but, 
in order to make them efficient, 
one must sample a different probability distribution, and thus one may be 
worried by the introduced bias.

In this paper we wish to discuss a family of dynamics that use 
bilocal moves and generalize the reptation dynamics: 
a bilocal move alters at the same time two
disjoint small groups of consecutive sites of the walk 
that may be very far away. 
Since a small number of beads is changed at each step, these algorithms
should be efficient in the presence of interactions, 
and thus they can be used in the study of the collapsed phase 
and of the folding of heteropolymers. 
Similar moves were introduced in 
Refs. \cite{Reiter_90,Jagodzinski-etal_92,Skolnick-Kolinski_91,foot2}.

The ergodicity properties of these algorithms have been discussed in 
Ref. \cite{Caracciolo-etal_2000}. Here, we will study the dynamic properties 
of two different implementations. The first one, the extended end-end
reptation (EER) algorithm, is obtained by performing at the same time 
reptation moves and bilocal kink-kink moves. Such an algorithm is quite 
efficient. In the absence of interactions, the autocorrelation time for 
quantities that measure the size of the walk---for instance, 
the end-to-end distance or 
the radius of gyration---scales as $N^z$ with $z\approx 2.2$ in 
two dimensions and 
$z\approx 2.1$ in three dimensions. The behavior of the energy 
(number of nearest-neighbor contacts among nonconsecutive links) is even faster,
with $z\approx 1.7$ in both dimensions. We have also tested the behavior of the algorithm at the $\theta$ point in two dimensions. We find that the critical 
behavior is only marginally worse, with $z\approx 2.3$, both for metric 
quantities and for the energy. We have also studied a different version,
the extended kink-end reptation (KER) algorithm, in which the reptation
moves are replaced by kink-end moves, i.e. by moves in which a kink is cleaved 
and two additional links are attached at the end of the walk and viceversa. 
This version is much slower, with $z\approx 2.9$ for quantities that measure 
the polymer size. Clearly, reptation moves 
are essential to obtain a fast dynamics.

The paper is organized as follows.
In Sec. \ref{sec2} we define the model and the observables whose critical 
behavior will be studied. In Sec. \ref{sec3} we define the basic moves 
and the two dynamics. Specific implementation details are reported in the 
Appendix. In Sec.~\ref{sec4} we define the autocorrelation times, the 
dynamic critical exponents, and the methods we use to compute them. 
In Secs.~\ref{sec5} and \ref{sec6} we discuss the critical behavior 
of the EER and of the KER algorithms in the absence of interactions,
while in Sec.~\ref{sec7} we discuss the behavior of the EER algorithm 
in two dimensions at the $\theta$ point. Conclusions are presented in 
Sec.~\ref{sec8}.

\section{Definitions} \label{sec2}

In this paper we consider SAWs with fixed number of steps $N$ 
and free endpoints on a hypercubic lattice $\mathbb{Z}^d$.
An $N$-step  SAW $\omega$ is a set 
of $N+1$ lattice sites $\omega_0$, $\ldots$, 
$\omega_N$, such that $\omega_i$ and $\omega_{i+1}$ are lattice 
nearest neighbors. By translation invariance we may fix $\omega_0$ to 
be the origin. 

We consider several observables
that measure of the size of an $N$-step SAW:
\begin{itemize}
\item The mean square end-to-end distance
\be
\label{raggioendtoend}
R_e^2 \equiv (\omega_N - \omega_0)^2\; .
\ee
\item The mean square radius of gyration
\be
\label{raggiogirazione} 
R_g^2 \equiv \frac{1}{N+1}\sum_{i=0}^{N}\Bigg(\omega_i -
\frac{1}{N+1}\sum_{k=0}^N \omega_k\Bigg)^2 =
\frac{1}{2(N+1)^2}\sum_{i,j=0}^{N}(\omega_i - \omega_j)^2 \; .
\ee
\item The mean square monomer distance from an endpoint
\be
\label{raggioquadra}
R_m^2 \equiv \frac{1}{N+1}\sum_{i=0}^{N}(\omega_i-\omega_0)^2 \; .
\ee
\end{itemize}
Moreover, for each SAW, we define the number of nearest-neighbor contacts
${\cal E}$. It is defined as follows. For every $i\not= j$ we define
\be
c_{ij} \equiv \cases{1 & $\qquad$ if $|\omega_i - \omega_j| = 1$; \cr
                0 & $\qquad$ otherwise.
               }
\ee
Then,
\be
{\cal E} \equiv - \sum_{i=0}^{N-2} \sum_{j=i+2}^N c_{ij}.
\ee
Note that we do not include here the trivial contacts between consecutive
walk sites.

We give each walk $\omega$ a weight $W(\beta)$ depending on the inverse
temperature $\beta \equiv 1/kT$,
\be 
W(\beta) \equiv {1\over Z_N} e^{-\beta {\cal E}},
\label{Gibbs-measure}
\ee
where 
$Z_N$ is the partition sum
\be 
Z_N = \sum_{\{\omega\}} e^{-\beta {\cal E}},
\ee
and the sum is extended over all $N$-step SAWs. 
\begin{table}[tbp]
\begin{center}
\begin{tabular}{lccc}
\hline\hline
\multicolumn{4}{c}{square lattice} \\
\hline
& year & method & $\beta_\theta$ \\
\hline
Ref. \protect\cite{Ishinabe_87} & 1987 & EE & 0.75 \\
Ref. \protect\cite{SS-88}       & 1988 & MC & $0.65 \pm 0.03$ \\
Ref. \protect\cite{MV-90}       & 1990 & EE & $0.67 \pm 0.04$ \\
Ref. \protect\cite{ML-90}       & 1990 & MC & $0.658 \pm 0.004$ \\
Ref. \protect\cite{FOT-92}      & 1992 & EE & $0.657 \pm 0.016$ \\
Ref. \protect\cite{CM-93}       & 1993 & MC & $0.658 \pm 0.004$ \\
Ref. \protect\cite{OPBG-94}     & 1994 & EE & $0.660 \pm 0.005$ \\
Ref. \protect\cite{GH-95}       & 1995 & MC & $0.665 \pm 0.002$ \\
Ref. \protect\cite{Nidras_96}   & 1996 & MC & 0.664, 0.666 \\
\hline
\hline
\multicolumn{4}{c}{cubic lattice} \\
\hline
& year & method & $\beta_\theta$ \\
\hline
Ref. \protect\cite{ML-89}     & 1989 & MC & $0.274 \pm 0.006 $\\
Ref. \protect\cite{GH-95b}    & 1995 & MC & $0.2687 \pm 0.0005$ \\
Ref. \protect\cite{TROW-96}   & 1996 & MC & $0.2779 \pm 0.0041$ \\
Ref. \protect\cite{TROW-96}   & 1996 & MC & $0.2782 \pm 0.0070$ \\
Ref. \protect\cite{TROW-96b}  & 1996 & MC & $0.276 \pm 0.006$ \\
\hline\hline
\end{tabular}
\end{center}
\caption{Estimates of $\beta_\theta$ on the square and cubic lattices. 
EE stands for exact enumeration, MC for Monte Carlo. 
}
\label{beta-theta}
\end{table}
The mean values $\<R_e^2\>_N$, $\<R_g^2\>_N$,  and $\<R_m^2\>_N$ 
have the asymptotic behavior 
\be
\label{ragginu}
\<R_e^2\>_N,\<R_g^2\>_N,\<R_m^2\>_N \sim N^{2\nu}
\ee   
as $N\rightarrow\infty$, where $\nu$ is a critical
exponent, which is independent of the microscopic details of the 
lattice model but depends on $\beta$. For $\beta < \beta_{\theta}$
(good-solvent regime) the SAW is swollen and  
(see Ref. \cite{PV_00} for a review of estimates of $\nu$ in three dimensions)
\be
\nu = \cases{\displaystyle{3\over4} & $\qquad$ $d = 2$, \cr
             \vphantom{\displaystyle{3\over4}} 
                          0.58758(7) & $\qquad$ $d = 3$ \ (Ref. \cite{BN-97}),
            }
\ee
while for $\beta > \beta_\theta$ the SAW is compact with 
\be 
\nu = {1\over d}\; .
\ee
For the very specific value $\beta = \beta_\theta$ ($\theta$-point) 
\be
\nu = \cases{\displaystyle{4\over7} & $\qquad$ $d = 2$\ 
       (Ref. \cite{saleurdupl}); \cr
       \vphantom{,} \cr
       \displaystyle{1\over2}\times{\rm log} & 
                                        $\qquad$ $d = 3$\ 
(Refs. 
   \cite{deGennes_book,Stephen_75,Duplantier_82,Duplantier_86,Duplantier_87}),
            }
\ee
The critical value $\beta_\theta$ depends on the model, on the lattice type,
and on all microscopic details. For the model we consider here on the square 
and on the cubic lattice, the best estimates are reported in Table 
\ref{beta-theta}.

\section{Algorithms} \label{sec3}

For the simulation of weakly interacting walks, i.e. for 
$\beta < \beta_\theta$, there exist powerful nonlocal algorithms
\cite{Madras-Sokal_88,Kennedy_01}.
However, these algorithms cannot be used in confined geometries---they
are not ergodic---and are very inefficient in the presence of strong 
interactions. Indeed, in these conditions nonlocal moves are rarely
accepted. In this paper we consider two algorithms that use 
local and bilocal moves \cite{Caracciolo-etal_2000}. 
A local move is one that 
alters only a few consecutive beads of the SAW, leaving the other sites 
unchanged. A bilocal move is instead one that alters two disjoint small groups 
of consecutive sites of the walk; these two groups may in general be very
far from each other. 

\begin{figure}
\centering
\epsfig{figure=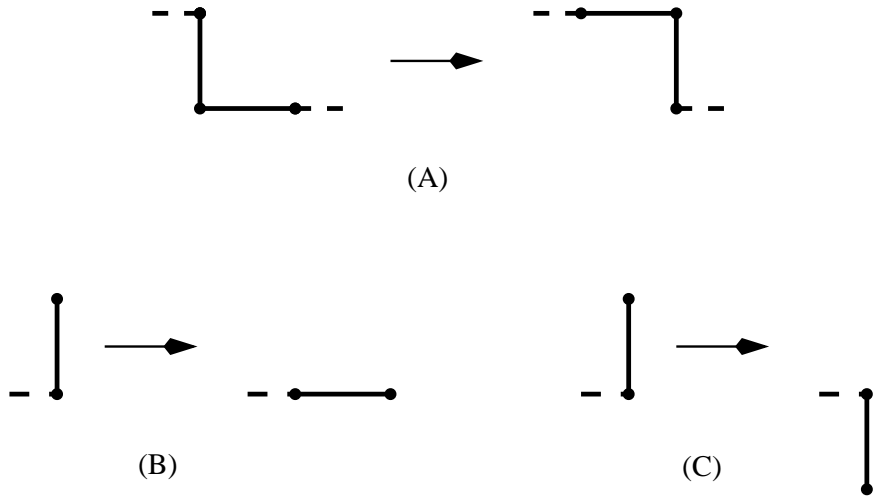,angle=-90,width=0.7\linewidth}
\vspace{0.5cm}
\caption{All one-bead moves: (A) One-bead flip. (B) $90^{\circ}$
end-bond rotation. (C) $180^{\circ}$ end-bond rotation.}
\label{one-bead}
\end{figure}


In our study we consider two types of local moves 
(see Fig. \ref{one-bead}):
\begin{itemize}
\item[{[L0]\hphantom{0}}] One-bead flips in which one {\em internal} 
bead (i.e. $\omega(i)$, $1\le i\le N-1$) only is moved.
\item[{[L1]\hphantom{0}}] End-bond rotations in which the last step of the 
walk is rotated. 
\end{itemize}

\begin{figure}
\centering
\epsfig{figure=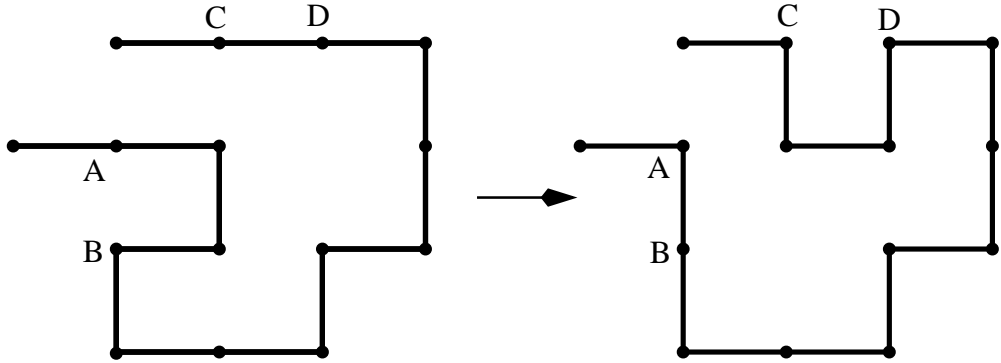,angle=-90,width=0.8\linewidth}
\vspace{0.5cm}
\caption{The kink-transport move. A kink has been cleaved from AB and
attached at CD. Note that the new kink is permitted to occupy one or
both of the sites abandoned by the old kink.}
\label{kink-trans}
\end{figure}

\begin{figure}
\centering
\vspace{0.2cm}
\epsfig{figure=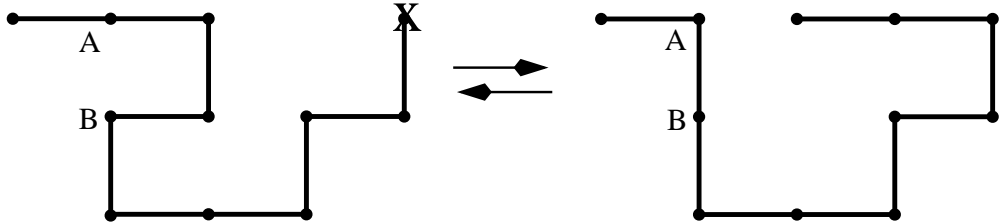,angle=-90,width=0.8\linewidth}
\vspace{0.5cm}
\caption{The kink-end reptation $(\longrightarrow)$ and end-kink
reptation $(\longleftarrow)$ moves. In $(\longrightarrow)$, a kink has
been cleaved from AB and two new steps have been attached at the end
marked X. Note that the new end steps are permitted to occupy one or
both of the sites abandoned by the kink.}
\label{kink-end}
\end{figure}

\begin{figure}
\centering
\epsfig{figure=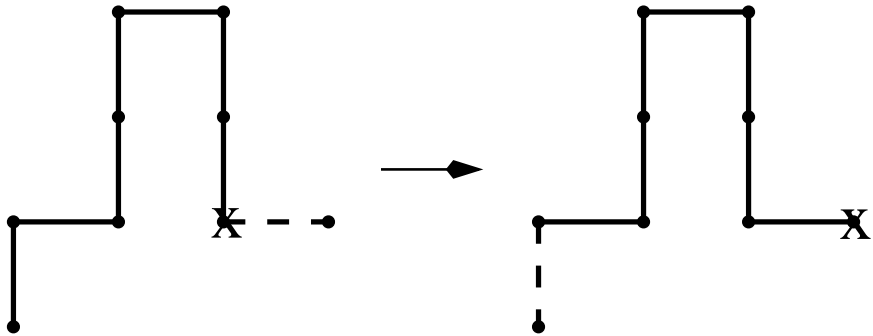,angle=-90,width=0.7\linewidth}
\vspace{0.5cm}
\caption{The reptation move. The head of the walk is indicated by
X. The dashed lines indicate the proposed new step and the 
abandoned old step.}
\label{reptation}
\end{figure}

We also introduce several types of bilocal moves:
\begin{itemize}
\item[{[B22]}] Kink-transport moves in which a kink is cleaved from
the walk and attached at a pair of neighboring sites somewhere else 
along the walk (see Fig. \ref{kink-trans});
note that the new kink is allowed to occupy one or both of the sites 
abandoned by the old kink.
\item[{[BKE]}] Kink-end and end-kink reptation moves (see Fig. \ref{kink-end}).
In the kink-end reptation move a kink is deleted at one location along the 
walk and two new bonds are appended in arbitrary directions at the free 
endpoint of the walk. Viceversa, an end-kink reptation move consists in 
deleting two bonds from the end of the walk and in inserting a kink, 
in arbitrary orientation, at some location along the walk.
\item[{[BEE]}] Reptation move (see Fig. \ref{reptation}) in which one bond is 
deleted from one end of the walk and a new bond is appended in arbitrary 
direction at the other end. 
\end{itemize}

Using these elementary moves we introduce three different updates that leave 
invariant the Gibbs measure \reff{Gibbs-measure}. They are described 
in detail in the Appendix. We will then consider two algorithms:
\begin{itemize}
\item[(a)] Extended end-end reptation (EER) algorithm;
\item[(b)] Extended kink-end reptation (KER) algorithm.
\end{itemize}

The EER algorithm consists in combining with 
non-zero probability the reptation move and the kink-kink local/bilocal
move, see Appendices \ref{App.A.1} and \ref{App.A.2} for the 
implementation details.  More precisely, 
with probability $p$ one performs a reptation move, 
and with probability $1-p$ a kink-kink local/bilocal move.
The KER algorithm works analogously: instead of the reptation moves
we use the kink-end/end-kink BKE moves. 
Both algorithms are known to be ergodic in 
two dimensions, while for $d=3$ ergodicity has been proved only for 
the KER algorithm \cite{Caracciolo-etal_2000}. For the EER algorithm 
ergodicity is still an open problem.
The probability $p$ is not fixed, and can be tuned to obtain 
the fastest dynamics. However, it should not have any
influence on the dynamic critical behavior, i.e. on the dynamic critical 
exponents. For this reason we have not performed a systematic 
study of the dependence of the numerical results on $p$ and we have 
simply set $p=0.50$ in most of our simulations. 

In order to have a fast and efficient implementation, it is important that
local and bilocal moves are performed in a CPU time of order one, 
i.e. constant as $N\to\infty$. This can be obtained only with a careful choice 
of the data structures used to store the walk coordinates. 
For local and end-end reptation moves it is sufficient to store the 
walk coordinates as a circular list in which the coordinates of the beads 
are stored sequentially. However, such a data structure is not 
convenient for B22 and BKE moves, since in this case insertion and deletion of
a single point requires a time of order $N$. The problem is solved 
\cite{Sokal_review} by 
storing the coordinates in a {\em contiguously allocated doubly-linked 
linear list}. The coordinates are stored contiguously but not in any 
particular order. In order to find the bead that follows and precedes a 
given one, one keeps two arrays of pointers that give the location 
in the coordinate list of the preceding and of the successive walk bead.
With this type of data structure, insertions and deletions take a time 
of order one. An efficient implementation requires also the ability 
to perform the self-avoidance check in a time of order one. 
This may be obtained by using a bit table---however, this is only possible 
for short walks---or a hash table. In this second case, one must 
be careful to use a hashing method that allows to insert and delete a 
single point in a time of order one. A very efficient method 
is described in Ref. \cite{Knuth}.
 
\section{Dynamical behavior} \label{sec4}

In order to determine the efficiency of the dynamics, we study the 
critical behavior of the autocorrelation times. 
Quite generally, given an observable $A$, we define
the unnormalized autocorrelation function
\be
C_{AA}(t) \equiv  \<A_s A_{s+t}\> - \<A\>^2,   
\ee
and its normalized counterpart
\be
\rho_{AA} \equiv C_{AA}(t)/C_{AA}(0),
\ee
where $t$ is the dynamics ``time."
Typically, $\rho_{AA}(t)$ decays exponentially, i.e. 
$\rho_{AA}(t) \sim e^{-|t|/\tau}$, for
large $|t|$. Thus, we can define the {\em exponential} autocorrelation time
\be
\tau_{{\rm exp},A} = \limsup_{|t| \rightarrow +\infty}
\frac{|t|}{-\ln|\rho_{AA}(t)|}
\label{def-tauexp}
\ee
and
\be
\tau_{\rm exp} = \sup_A \tau_{{\rm exp},A}.
\ee
Thus, $\tau_{\rm exp}$ is the relaxation time of the slowest mode in the
system. 

We also define the {\em integrated} autocorrelation time
\be 
\tau_{{\rm int},A} = \frac{1}{2} \sum_{s=-\infty}^{+\infty} \rho_{AA}(s) = 
    \frac{1}{2} + \sum_{s=1}^{+\infty} \rho_{AA}(s),
\ee 
which controls the
statistical error in Monte Carlo measurements of $\<A\>$. 
The factor $1/2$ is inserted so that 
$\tau_{{\rm int},A}\approx\tau_{{\rm exp},A}$ if
$\rho_{AA}(t)= e^{-|t|/\tau_{{\rm exp},A}}$ with
$\tau_{{\rm exp},A} \gg 1$. 

In order to estimate $\tau_{{\rm exp},A}$, we will proceed as follows.
If $\hat{\rho}_{AA}(t)$ is the normalized autocorrelation
function estimated from the data, 
we define an effective exponent
\be
\hat{\tau}_{{\rm exp},A} (t;s) \equiv 
    s\left[ \ln \left( { \hat{\rho}_{AA}(t)\over \hat{\rho}_{AA}(t+s)}\right) \right]^{-1}\; ,
\ee
where $s$ is some fixed number. This quantity should become independent 
of $t$ as $t\to\infty$. In practice, we look for a region in which the 
estimates are sufficiently stable with $t$ and then take the 
value of $\hat{\tau}_{{\rm exp},A} (t;s)$ in this region as our
estimate of $\tau_{{\rm exp},A}$. In order to apply the method we should 
also choose the parameter $s$. It should be neither too small, 
i.e. $s\ll \tau_{{\rm exp},A}$, otherwise 
$\hat{\rho}_{AA}(t)\approx \hat{\rho}_{AA}(t+s)$, nor too large, 
i.e. $s\gg \tau_{{\rm exp},A}$, otherwise the error on 
$\hat{\rho}_{AA}(t+s)$ is large. In our study we have always 
fixed $s$ self-consistently, by taking 
$s\approx \tau_{{\rm exp},A}/k$ with $k\approx10$-20.

In order to estimate $\tau_{{\rm int},A}$, we use the 
self-consistent windowing method proposed in Ref. \cite{Madras-Sokal_87}.
We define our estimate as
\be 
\hat{\tau}_{{\rm int},A} = \frac{1}{2} \sum_{t=-M}^{M} \hat{\rho}_{AA}(t),
\label{tauint-automatic-win}
\ee  
where $M$ is the smallest integer that satisfies $M\ge c\, \hat{\tau}_{{\rm int},A}$
and $c$ is a fixed constant. 
The variance of $\hat{\tau}_{{\rm int},A}$ is given by
\be 
\label{vartau} 
\textrm{var}(\hat{\tau}_{{\rm int},A}) \approx 
\frac{2(2M +1)}{n}\hat{\tau}^2_{{\rm int},A} ,
\ee 
where $n$ is the number of Monte Carlo iterations and we have made 
the approximation $\tau_{{\rm int},A} \ll M \ll n$. 
The method provides quite accurate estimates of $\hat{\tau}_{{\rm int},A}$ 
as long as $c$ is chosen so that $M$ is a few times $\tau_{{\rm exp},A}$. 
There are cases in which $\tau_{{\rm int},A} \ll \tau_{{\rm exp},A}$, 
so that the previous condition is difficult to satisfy. 
In this case, a successful ``ad hoc'' recipe was proposed by 
Li \emph{et al.}~\cite{limadsok} for the {pivot}
algorithm. They noted that, when $A$ is one of the radii,
$\rho_{AA}(t)\approx 1/t^q$ in the intermediate region
$\tau_{{\rm int},A} \ll t \lesssim\tau_{{\rm exp},A}$, 
with $q\approx 1$-1.2. Thus, 
they 
extrapolated $\rho_{AA}(t)$ proportionally to $1/t$ for $t > M$,
(i.e. set ${\hat{\rho}_{AA}}(t) = M{\hat{\rho}_{AA}}(M)/t$ for $t > M$),
and then used this expression to compute the contribution to 
$\tau_{{\rm int},A}$ from the region $M<t<\tau_{{\rm exp},A}$. 
This gives the modified 
estimate
\be 
\tilde{\tau}_{{\rm int},A} = \frac{1}{2} \sum_{t=-M}^{+M} 
      {\hat{\rho}_{AA}}(t) + 
M \hat{\rho}_{AA}(M)\ln\bigg(\frac{\tau_{{\rm exp},A}}{M}\bigg). 
\label{tauint-Lietal}
\ee 
In general, the autocorrelation times diverge as $N\to \infty$. 
We can thus define two dynamic critical exponents 
$z_{{\rm exp},A}$ and $z_{{\rm int},A}$ by
\bea
\tau_{{\rm int},A} &\sim & N^{z_{{\rm int},A}}, \nonumber \\
\tau_{{\rm exp},A} &\sim & N^{z_{{\rm exp},A}}. \label{dynamic-scaling-form}
\eea
It is important to stress that in general 
$z_{{\rm int},A} \not= z_{{\rm exp},A}$ and that 
different observables may have different critical exponents.

The dynamic critical exponents can also be obtained by requiring the 
autocorrelation function to obey a dynamic 
scaling law of the form~\cite{dinamscal,Sokal_review,Caracciolo-etal_90}
\be 
\label{scaling_law1} 
\rho_{AA}(t;N) \approx |t|^{-a}F_{AA}\left(\,t\,N^{-b}\right)  \approx
                        N^{-ab} G_{AA}\left(\,t\,N^{-b}\right),
\ee
valid in the limit $N\to\infty$,  $|t|\to \infty$, with 
$t\,N^{-b}$ fixed. 
Here, $a,b>0$ are dynamic critical exponents and $F_{AA}$ and 
$G_{AA}$ are  suitable scaling functions.
If $F_{AA}(x)$ is continuous,
strictly positive, and rapidly decaying---e.g. exponentially---as 
$|x|\rightarrow\infty$, then, it is not hard to
see that  
\bea 
\label{scale_pred1} 
z_{{\rm exp},A}\ &= & b,\\
\label{scale_pred2}
z_{{\rm int},A}\ &= & (1-a)b \quad 
(\textrm{if}\ a\,<\,1\,)\; .
\eea 

The exponents $a$ and $b$ can be determined by requiring the collapse onto a 
single curve of the autocorrelation functions corresponding to 
different values of $N$. Then, using the previous formulae, 
one can determine $z_{{\rm exp},A}$ and $z_{{\rm int},A}$.
The advantage of this method is in its 
bypassing the problem of determining $\tau_{{\rm exp},A}$ and 
$\tau_{{\rm int},A}$,
but it is quite difficult to assess the errors, since the optimal 
values are determined visually.

\section{The EER dynamics in two and three dimensions}  \label{sec5}
 
We performed an extensive Monte Carlo simulation of 
noninteracting ($\beta = 0$) SAWs on the 
square lattice, $d=2$,
and on the cubic lattice, $d=3$, using the EER algorithm. 
We set $p=0.5$ and used the first version of the reptation
move, see App.~\ref{App.A.2}.
We considered only three  values of $N$, $N=100,300,1000$, 
but for each of them we collected a very large statistics, 
see Table \ref{rept_tauexp}. 
We measured
$\< R_g^2 \>_N$, $\< R_e^2 \>_N$, $\< R_m^2 \>_N$, and the 
energy $\mathcal{E}_N$, i.e. the number of nearest-neighbor
contacts.
We compared the static results of our simulations with those of 
Li \emph{et al.}~\cite{limadsok}, finding good agreement.

\begin{figure}[!t]
\begin{center}
\epsfig{figure=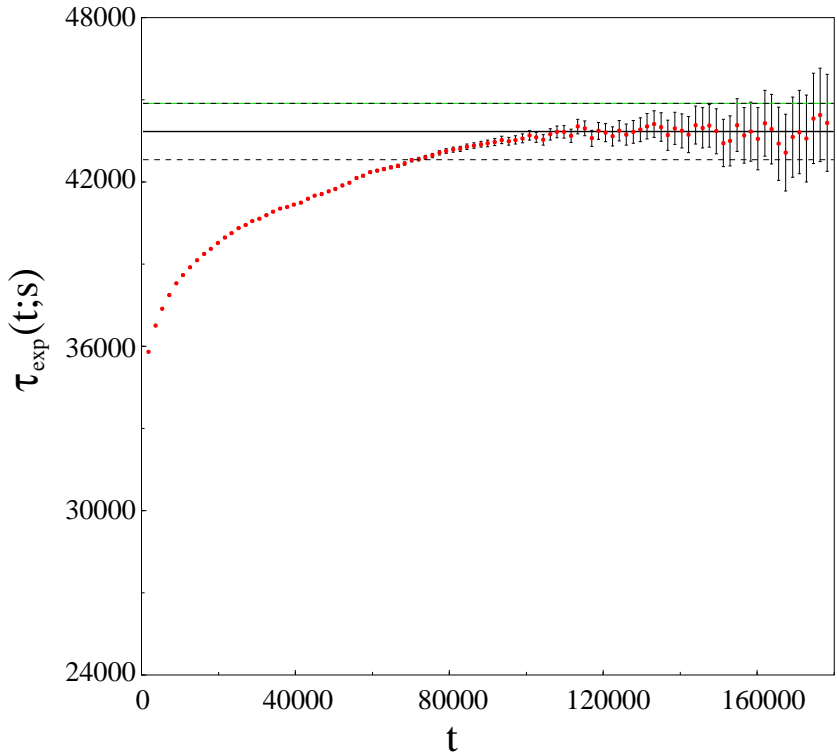,angle=0,width=0.7\linewidth}
\end{center}
\caption{Effective exponent $\hat{\tau}_{{\rm exp},R^2_g}(t;s)$ for the 
EER algorithm. Here $d=2$, $N=300$, $s = 2000$. 
The horizontal line corresponds to our final estimate 
${\tau}_{{\rm exp},R^2_g} = 43600$ and the dashed lines to 
$\pm$ one error bar (900).
}
\label{effmass-reptation}
\end{figure}

First of all, we measured the exponential autocorrelation times.
As an example, in Fig. \ref{effmass-reptation} we report 
the effective exponent
$\hat{\tau}_{{\rm exp},R^2_g}(t;s)$ for $N=300$, $d=2$, $s = 2000$. 
It increases with $t$ and becomes constant within error bars 
for $t \gtrsim 106000$.  We have taken our estimate at 
$t \approx 130000 \approx 3 {\tau}_{{\rm exp},R^2_g}$. 
The same analysis has been done for 
all observables and all values of $N$. The results are reported in 
Table \ref{rept_tauexp}.

\begin{table}[!t]
\protect\footnotesize
\centering
\begin{tabular}{cccr@{$\,\pm\,$}lr@{$\,\pm\,$}lr@{$\,\pm\,$}lr@{$\,\pm\,$}l}
\hline\hline
$d$ & $N$ & iter & \multicolumn{2}{c}{\large{$\tau_{{\rm exp},R_g^2}$}} &
\multicolumn{2}{c}{\large{$\tau_{{\rm exp},R_e^2}$}} &
\multicolumn{2}{c}{\large{$\tau_{{\rm exp},R_m^2}$}} &
\multicolumn{2}{c}{\large{$\tau_{{\rm exp},\cal E}$}} \\
\hline
  &100 & $5.77\cdot 10^{10}$ &
      3230 & 20 & 3062 & 24 & 3190 & 16 & 3078 & 40 \\
2 &300 & $5.38\cdot 10^{11}$ &
      43600 & 900 & 47300 & 1500 & 43140 & 760 & 35600 & 1260 \\
  &1000 &$6.90\cdot 10^{11}$ & 
     710400 & 42000 & 859600 & 46000 & 745000 & 36000 & 448000 & 30000 \\
\hline
  &100 & $4.36\cdot 10^{10}$ &
     2980 & 40 & 3020 & 30 & 2968 & 30 & 2840 & 60 \\
3 &300 & $4.42\cdot 10^{11}$ &
     32640 & 840 & 34700 & 900 & 32980 & 840 & 27480 & 800\\
  &1000 & $5.20\cdot 10^{11}$&
    370400 & 22000 & 374800 & 24000 & 368000 & 22000 & 300800 & 16000\\
\hline\hline
\end{tabular}
\caption{Exponential autocorrelation times for the EER algorithm
in two and three dimensions. ``iter" is the number of iterations. }
\label{rept_tauexp}
\end{table}

\begin{table}[!t]
\protect\footnotesize
\centering
\begin{tabular}{ccr@{$\,\pm\,$}lr@{$\,\pm\,$}lr@{$\,\pm\,$}lr@{$\,\pm\,$}l}
\hline\hline
$d$ &  & \multicolumn{2}{c}{\large{$\tau_{{\rm exp},R_g^2}$}} &
\multicolumn{2}{c}{\large{$\tau_{{\rm exp},R_e^2}$}} &
\multicolumn{2}{c}{\large{$\tau_{{\rm exp},R_m^2}$}} &
\multicolumn{2}{c}{\large{$\tau_{{\rm exp},\cal E}$}} \\
\hline
&$z_{\rm exp}$ & 2.37 & 0.02 & 2.46 & 0.02 & 2.36 & 0.02 & 2.18 & 0.02\\
2 &$B$ & 
    0.0610 & 0.0048 & 0.0376 & 0.0032 & 0.0586 & 0.0030 & 0.133 & 0.013 \\
&$\chi^{2}$ & \multicolumn{2}{c}{1.18} & \multicolumn{2}{c}{1.03} &
\multicolumn{2}{c}{0.52} & \multicolumn{2}{c}{2.33}\\
\hline
&$z_{\rm exp}$ & 2.10 & 0.03 &
2.11 & 0.04 & 2.10 & 0.03 & 2.04 & 0.03 \\
3 &$B$ & 0.196 & 0.032 &
0.180 & 0.028 & 0.190 & 0.030 & 0.242 & 0.028 \\
&$\chi^{2}$ & \multicolumn{2}{c}{1.13} & \multicolumn{2}{c}{1.90} &
\multicolumn{2}{c}{1.31} & \multicolumn{2}{c}{1.57}\\
\hline\hline
\end{tabular}
\caption{Dynamic exponent $z_{\rm exp}$ for the EER algorithm in two and 
three dimensions, obtained by fitting $\tau_{\rm exp} = B N^{z_{\rm exp}}$.
The number of degrees of freedom of the fit is 1.}
\label{rept_tauexp_fit}
\end{table}

We have determined the dynamic exponent $z_{{\rm exp},A}$ by fitting the
results of Table \ref{rept_tauexp} with the Ansatz
\be
\tau_{{\rm exp},A} = B_A N^{z_{{\rm exp},A}}.
\label{Ansatzexp}
\ee
The results are reported in Table \ref{rept_tauexp_fit}. 
Since we only have three values of $N$, we cannot study the effect of 
corrections to scaling and thus we cannot determine the 
systematic error on our results. However, an indication can be obtained by 
comparing the results for the three radii. Indeed, one expects these 
three observables to have the same dynamic exponent. In two dimensions,
$z_{{\rm exp},R^2_e}$ differs significantly from the estimates for the 
other radii, indicating the presence of large corrections to scaling. 
Such conclusion is confirmed by the scaling analysis that will be 
reported below: as we shall see, the results of Table \ref{rept_tauexp_fit}
in two dimensions are effective exponents that are expected to decrease as 
larger values of $N$ are included in the fit. Of course, 
it is possible that also $z_{{\rm exp},\cal E}$ is affected by large 
corrections to scaling. However, the scaling analysis confirms the 
result for $z_{{\rm exp},\cal E}$, thus showing that the scaling 
corrections are important for the radii only.

In three dimensions, there is no evidence of large scaling corrections.
Indeed, the exponents $z_{\rm exp}$ for the three radii agree within error bars.
The dynamic exponent for the energy is lower than that for the radii. 
However, the difference is small and it is not clear if 
it is real or just the result of neglected corrections to scaling.

Then, we have computed the integrated autocorrelation times by using the 
self-consistent windowing method with $c=15$, 
cf. Eq. (\ref{tauint-automatic-win}). The results are reported
in Table \ref{rept_tauint}.

\begin{table}[!b] 
\protect\footnotesize 
\hspace{-0.4cm} 
\begin{center}
\begin{tabular}{ccr@{$\,\pm\,$}lr@{$\,\pm\,$}l
                       r@{$\,\pm\,$}lr@{$\,\pm\,$}l} 
\hline \hline
$d$ & $N$ & 
\multicolumn{2}{c}{\large{$\tau_{{\rm int},R_g^2}$}} & 
\multicolumn{2}{c}{\large{$\tau_{{\rm int},R_e^2}$}} & 
\multicolumn{2}{c}{\large{$\tau_{{\rm int},R_m^2}$}} & 
\multicolumn{2}{c}{\large{$\tau_{{\rm int},{\cal E}}$}} \\ 
\hline 
 &100 &  3113.6 & 5.6 & 2209.2 & 3.4 & 2124.4 & 3.2 &  861.64 & 0.82\\ 
2&300 &  36000 & 72 & 24250 & 40 & 23764 & 38 &  5596.4 & 4.4\\ 
&1000 &  522880 & 3520 & 337220 & 1820 & 334600 & 1800 & 39224 & 72 \\ 
\hline 
 &100 & 2729.0 & 5.2 & 1761.2 & 2.8 & 1780.0 & 2.8 & 925.4 & 1.0\\
3&300 & 27092 & 52 & 16772 & 26 & 17162 & 26 &  6035.8 & 5.4\\
&1000 & 332900 & 2100 & 199160 & 960 & 205100 & 1000 & 42954 & 96 \\
\hline  
\hline 
\end{tabular} 
\end{center}
\caption{Integrated autocorrelation times for the EER algorithm
in two and three dimensions.}
\label{rept_tauint} 
\end{table} 

For the radii, $\tau_{{\rm exp},R^2} \sim 1$-$2\, \tau_{{\rm int},R^2}$ 
and thus the 
choice $c=15$ should be rather conservative. For the energy, $c=15$ corresponds
to $1.3\, \tau_{{\rm exp},\mathcal E} \lesssim M \lesssim 4\, \tau_{{\rm
exp},\mathcal E}$ in two dimensions and to
$2\, \tau_{{\rm exp},\mathcal E} \lesssim M \lesssim 5\, \tau_{{\rm
exp},\mathcal E}$ in three dimensions for our values of $N$ ($M$ is 
the cutoff defined
in Eq. \reff{tauint-automatic-win}). Such values of $M$ 
can give rise, at least for $N=1000$, to a systematic underestimate of 
$\tau_{{\rm int},\mathcal E}$. For our choice of $c$ and for $N=1000$, 
$\rho_{\mathcal E \mathcal E}(M) \approx (7.2\pm 1.5)\cdot 10^{-4}$ and 
$(4.9\pm 1.0)\cdot 10^{-4}$ in two and three 
dimensions respectively. Therefore, assuming a pure exponential 
behavior for $t>M$, the neglected contribution 
should be of order $320\pm60$ in $d=2$ and $150\pm30$ in $d=3$.
The correction is quite small, but larger than the quoted error bars:
the correct errors are at least larger 
by a factor of 5 (resp. 2) in two (resp. three) dimensions. 

The results for $\tau_{{\rm int},A}$ have been fitted with the Ansatz 
\be
\tau_{{\rm int},A} = B_A N^{z_{{\rm int},A}}.
\label{Ansatzint}
\ee
The results are reported in Table \ref{rept_int_fit}.

\begin{table} 
\protect\tiny 
\centering 
\protect\footnotesize 
\begin{tabular}{cl
       r@{$\,\pm\,$}lr@{$\,\pm\,$}l
       r@{$\,\pm\,$}lr@{$\,\pm\,$}l
       r@{$\,\pm\,$}lr@{$\,\pm\,$}l
       r@{$\,\pm\,$}lr@{$\,\pm\,$}l} 
\hline \hline
$d$& & \multicolumn{2}{c}{\large{$\tau_{{\rm int},R_g^2}$}} & 
       \multicolumn{2}{c}{\large{$\tau_{{\rm int},R_e^2}$}} & 
       \multicolumn{2}{c}{\large{$\tau_{{\rm int},R_m^2}$}} & 
       \multicolumn{2}{c}{\large{$\tau_{{\rm int},{\cal E}}$}} \\ 
\hline 
&$z_{\rm int}$ & 
    2.227 & 0.002 & 2.182 & 0.002 & 2.198 & 0.002 & 1.673 & 0.001\\ 
2&$B$ & 0.1100 & 0.0010 & 0.0956 & 0.0008 & 0.0856 & 0.0008 & 0.392 & 0.002\\ 
&$\chi^{2}$ &  \multicolumn{2}{c}{0.660} & 
    \multicolumn{2}{c}{1.04} & \multicolumn{2}{c}{0.054} &
    \multicolumn{2}{c}{1470} \\ 
\hline  
&$z_{\rm int}$ & 
    2.088 & 0.002 & 2.052 & 0.002 & 2.062 & 0.002 & 1.681 & 0.001 \\ 
3&$B$ & 0.1820 & 0.0020 & 0.1386 & 0.0012 & 0.1338 & 0.0010 & 0.392 & 0.002 \\ 
&$\chi^{2}$ &  \multicolumn{2}{c}{0.804} & 
 \multicolumn{2}{c}{0.625} & \multicolumn{2}{c}{0.191} & 
 \multicolumn{2}{c}{851} \\ 
\hline  
\hline 
\end{tabular} 
\caption{Dynamic exponent $z_{\rm int}$ for the EER algorithm in two and 
three dimensions, obtained by fitting $\tau_{\rm int} = B N^{z_{\rm int}}$.
The number of degrees of freedom of the fit is 1.}
\label{rept_int_fit} 
\end{table} 
 
The error bars are purely statistical and, as in the case of $z_{\rm exp}$, 
large systematic errors may be present. We expect the quantities 
that measure the walk size
to have the same dynamic exponent $z_{\rm int}$. Thus, both in two and 
three dimensions, the reported errors are largely underestimated. 
More conservative estimates are
\bea 
z_{{\rm int},R^2} = 2.20 \pm 0.03 && \qquad \qquad d = 2, 
\nonumber \\
z_{{\rm int},R^2}  = 2.07 \pm 0.02 && \qquad \qquad d = 3.
\label{EER-zintR2}
\eea
The fit for the energy has a very large 
$\chi^2$. There are two reasons for this.
First, the statistical errors are underestimated, 
as we already discussed. Second, there may be large---compared to the tiny
statistical errors---corrections to scaling. For these reasons, the errors
quoted in Table \ref{rept_int_fit} for $z_{{\rm int}, \cal E}$ should not 
be taken seriously. A more realistic estimate is obtained by multiplying 
the errors by $\sqrt{\chi^2}$, which gives an error of $\pm 0.04$ 
on the exponent in both two and three dimensions. Such error is more realistic
and indeed is close to the error we quoted for the radii, cf. 
Eq.~\reff{EER-zintR2}.

\begin{table} 
\protect\footnotesize 
\centering 
\begin{tabular}{ccr@{$\,\pm\,$}lr@{$\,\pm\,$}lr@{$\,\pm\,$}lr@{$\,\pm\,$}l
                  r@{$\,\pm\,$}lr@{$\,\pm\,$}lr@{$\,\pm\,$}l} 
\hline \hline
 $d$&  & \multicolumn{2}{c}{$R_g^2$} & \multicolumn{2}{c}{$R_e^2$} & 
 \multicolumn{2}{c}{$R_m^2$} & \multicolumn{2}{c}{$\mathcal{E}$} \\ 
\hline 
2&$b$ & 2.21 & 0.02 & 2.20 & 0.02 & 2.22 & 0.02 & 2.14 & 0.02 \\ 
&$a$ & 0.0010 & 0.0010 & 0.0025 & 0.0020 & 0.002 & 0.002 &  0.285 & 0.025 \\ 
\hline  
3&$b$ & 2.085 & 0.015 & 2.068 & 0.015 & 2.085 & 0.015 & 2.030 & 0.025\\ 
 &$a$& 0.0012 & 0.0010 & 0.0025 & 0.0025 & 0.0015 & 0.0015 & 0.210 & 0.020\\ 
\hline  
\hline 
\end{tabular} 
\caption{Dynamic exponents $a$ and $b$ for the EER algorithm in two and three
dimensions.}
\label{rept_scaling_radii} 
\end{table} 

\begin{figure}
\hspace{-0.6cm}
\epsfig{figure=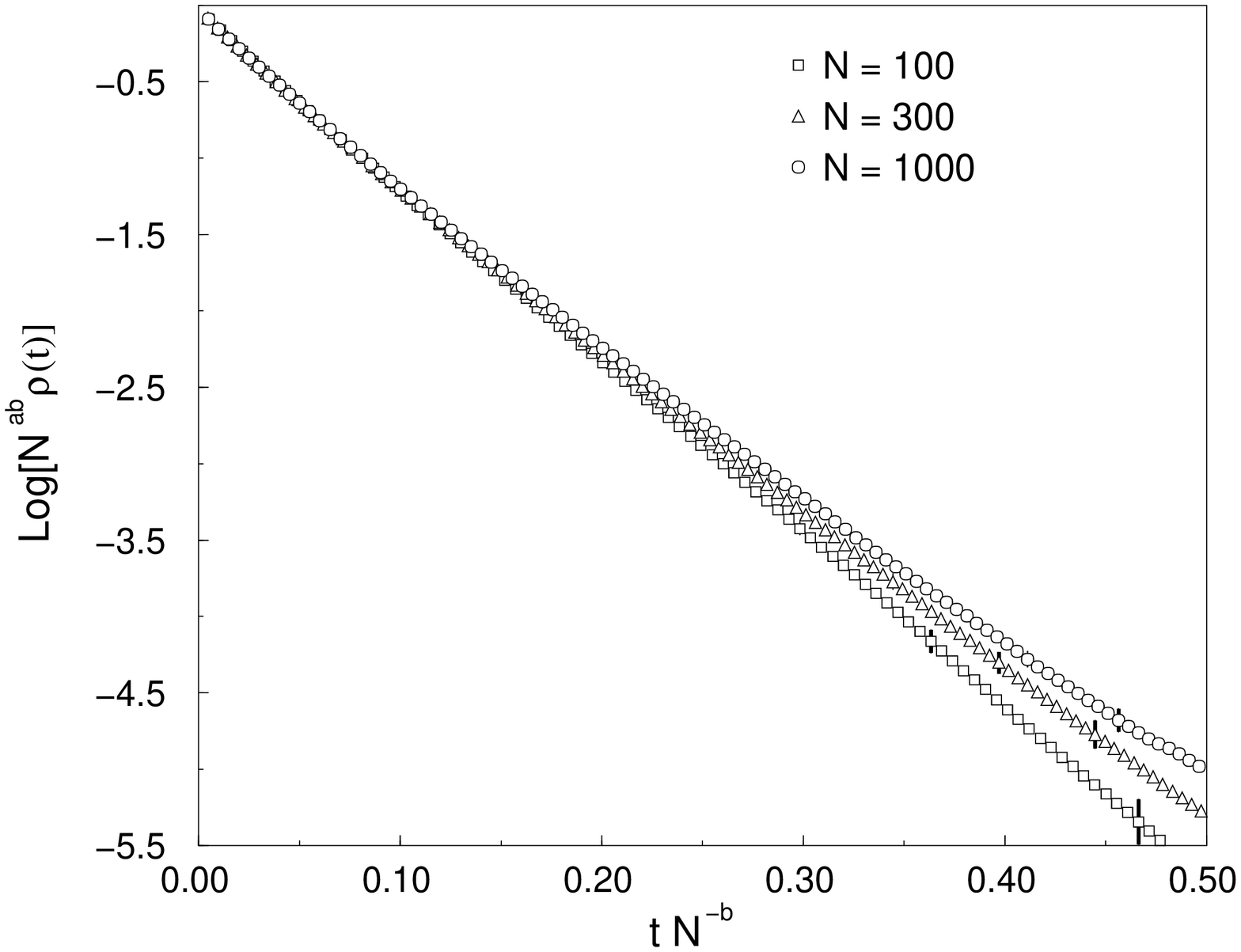,angle=-90,width=0.53\linewidth}
\epsfig{figure=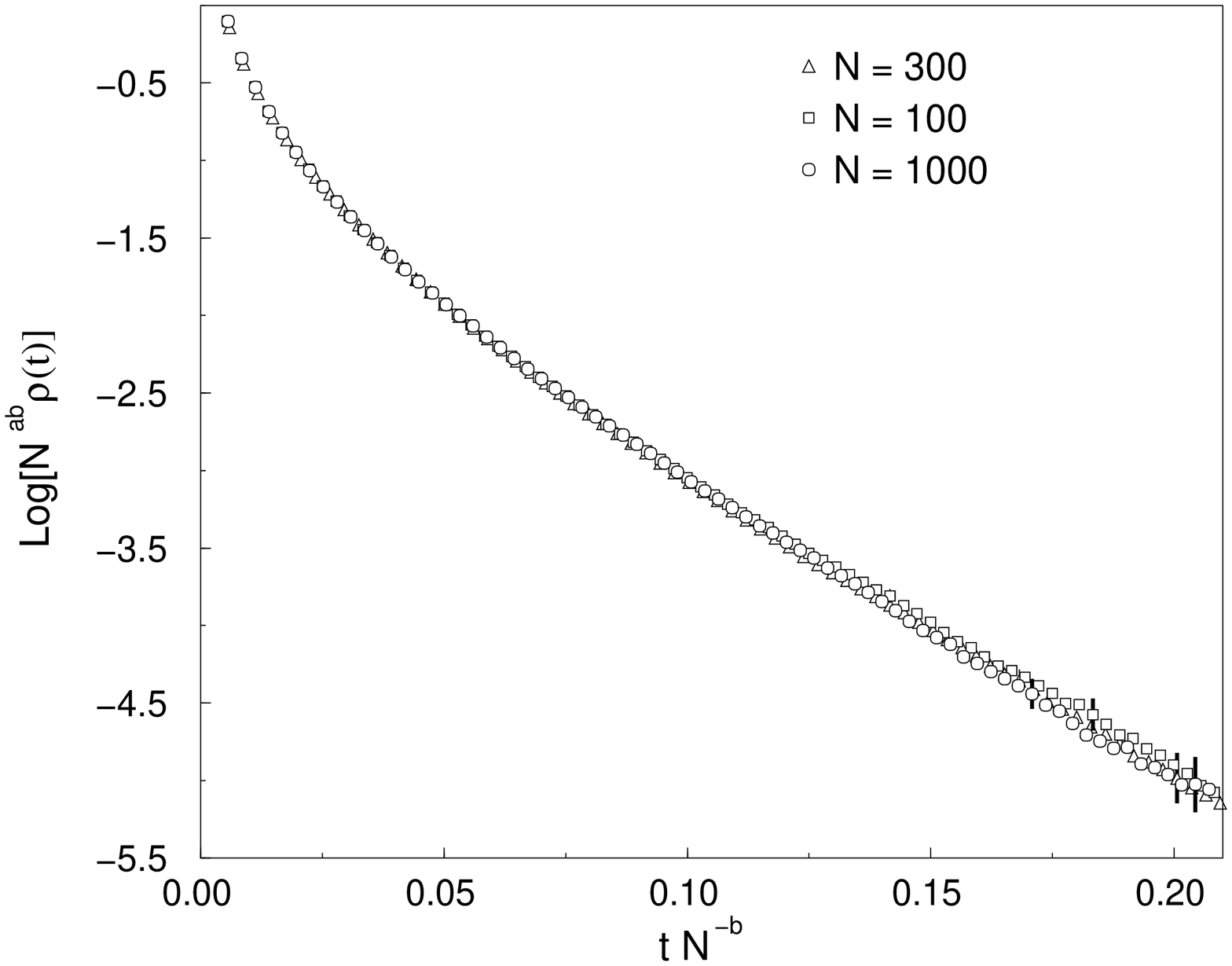,angle=-90,width=0.52\linewidth}
\caption{Dynamic scaling analysis for the EER algorithm in two dimensions: 
plots of 
$\ln[N^{ab} \hat{\rho}_{AA}(t)]$ vs. $tN^{-b}$. Left frame: 
$A = R^2_g$, $a = 0.001$, $b = 2.21$; 
right frame: $A = \mathcal E$, $a = 0.285$, $b = 2.14$.}
\label{dyn-scaling-rept-2d}
\end{figure}

The dynamic critical exponents can also be determined by analyzing 
the scaling behavior of the autocorrelation function, 
see Eq. \reff{scaling_law1}, i.e. by determining $a$ and $b$ 
so that $N^{ab} \hat{\rho}_{AA}(t)$ is a universal function of 
$t N^{-b}$, independent of $N$. 
In Table \ref{rept_scaling_radii} we report the values of $a$ and $b$ for which
a collapse of the autocorrelation functions is observed. 
In Fig. \ref{dyn-scaling-rept-2d} we show the corresponding plots for 
$R^2_g$ and $\cal E$ in two dimensions.

For the energy we observe a very good collapse, while for the radii the 
scaling behavior deteriorates as $t N^{-b}$ increases. Since the vertical scale
changes by several orders of magnitude, the deviations are somewhat difficult
to observe and this makes difficult to set the errors on $a$ and $b$. 
For this purpose, we found more useful to consider, instead of 
$N^{ab} \hat{\rho}_{AA}(t)$, the quantity
\be 
\tau_{{\rm scal},A} (t;N) = - {t N^{-b}\over \ln(N^{ab} \hat{\rho}_{AA}(t))},
\ee
that is also a universal function of $t N^{-b}$ in the scaling limit and 
that scales as $\tau_{{\rm exp},A} N^{-b}$ for $t N^{-b}\to \infty$. 

\begin{figure}
\hspace{-0.6cm}
\epsfig{figure=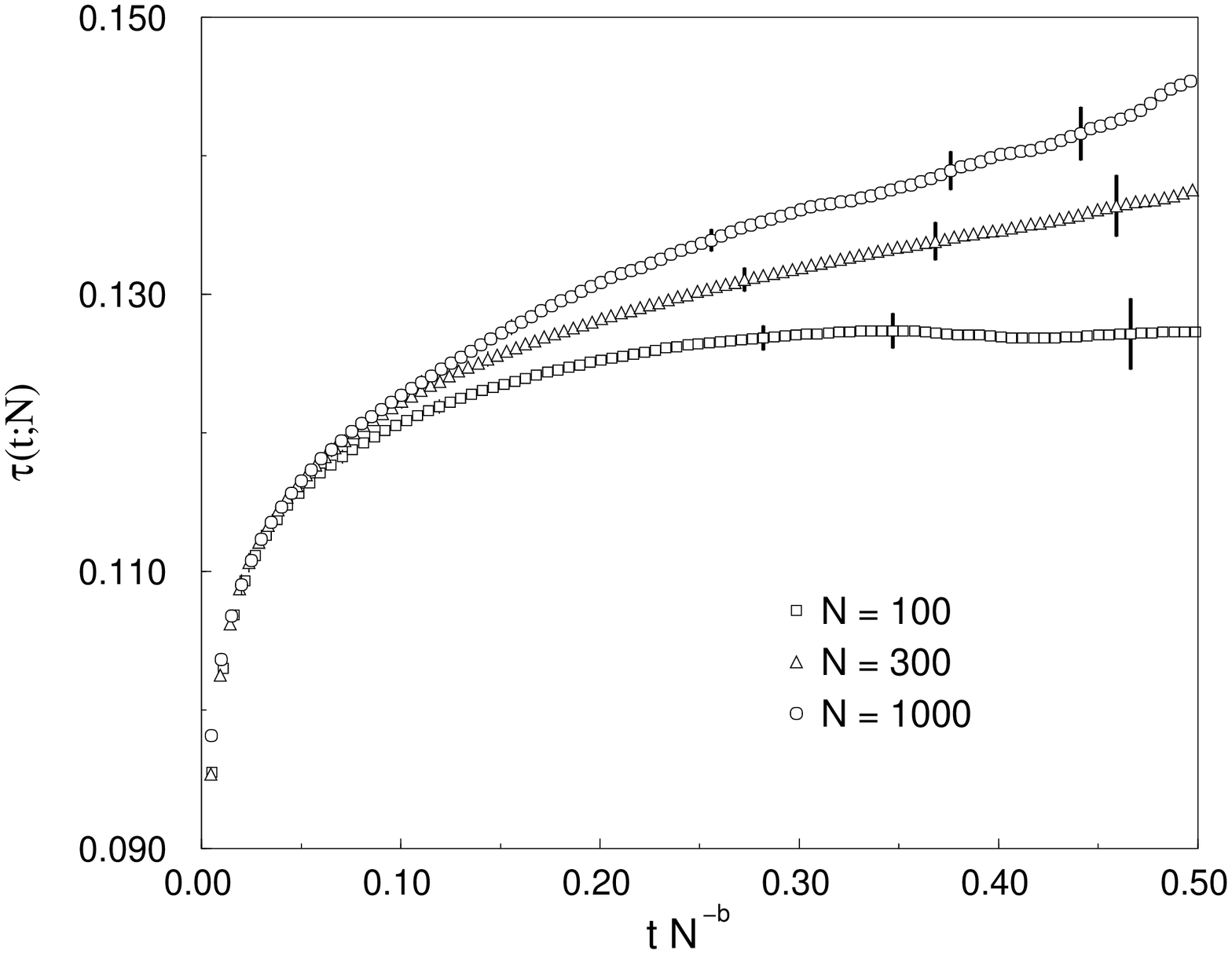,angle=-90,width=0.53\linewidth}
\epsfig{figure=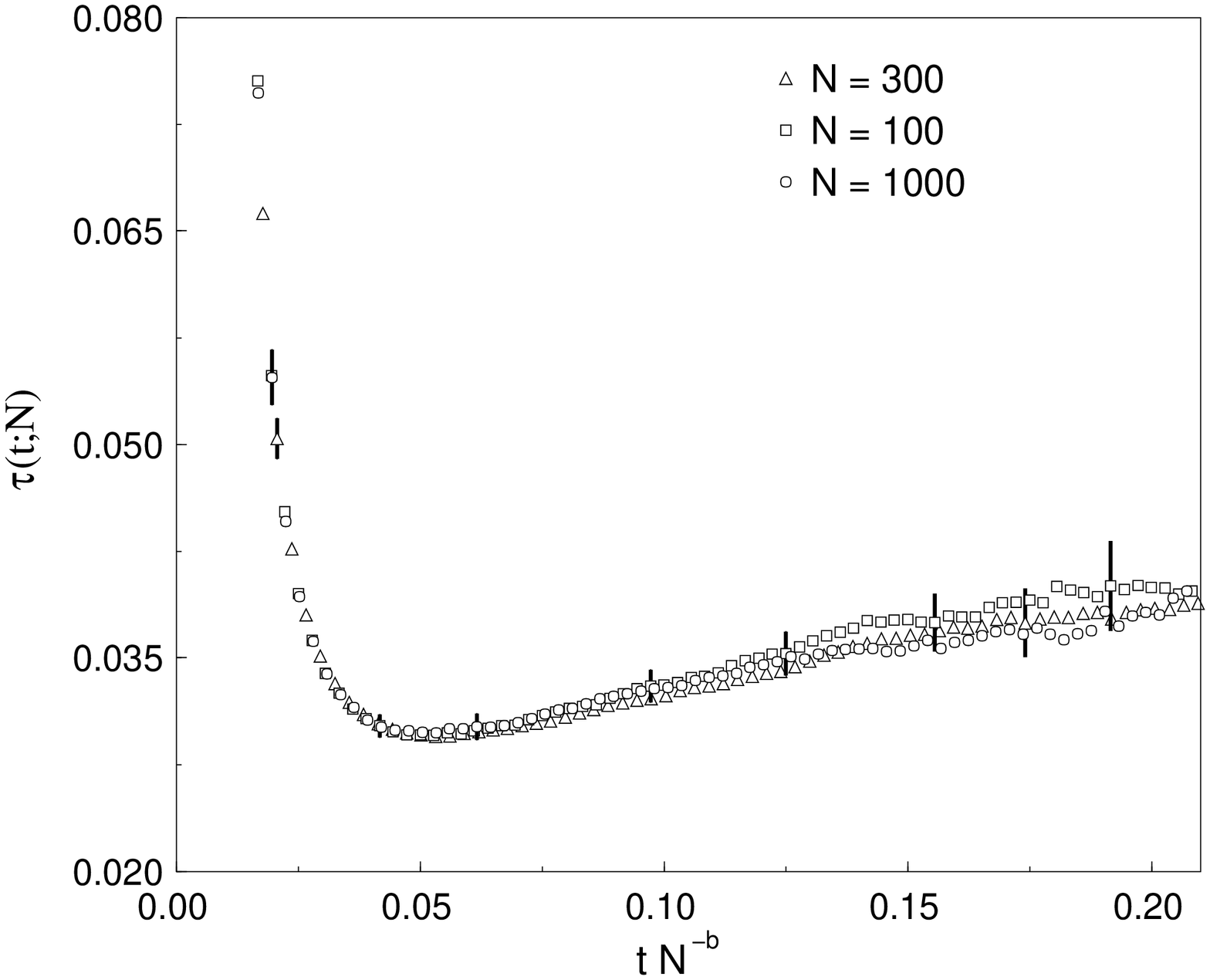,angle=-90,width=0.51\linewidth}
\caption{Dynamic scaling analysis for the EER algorithm in two dimensions: 
plots of 
$\tau_{{\rm scal},A}(t;N)$ vs. $tN^{-b}$. Left frame: 
$A = R^2_g$, $a = 0.001$, $b = 2.21$; 
right frame: $A = \mathcal E$, $a = 0.285$, $b = 2.14$.}
\label{dyn-scaling-rept-2d-bis}
\end{figure}

In Fig. \ref{dyn-scaling-rept-2d-bis} we report the quantity
$\tau_{{\rm scal},A}(t;N)$ for $R^2_g$ and $\mathcal E$. The
energy shows a very good scaling behavior while the scaling is quite poor 
for $R^2_g$, although it improves as $N$ increases: the data for 
$N=100$ and $N=300$ overlap up to $t N^{-b} \approx 0.05$,
while the data for $N=300$ and $N=1000$ overlap up to 
$t N^{-b} \approx 0.09$. 

\begin{figure}
\hspace{-0.6cm}
\epsfig{figure=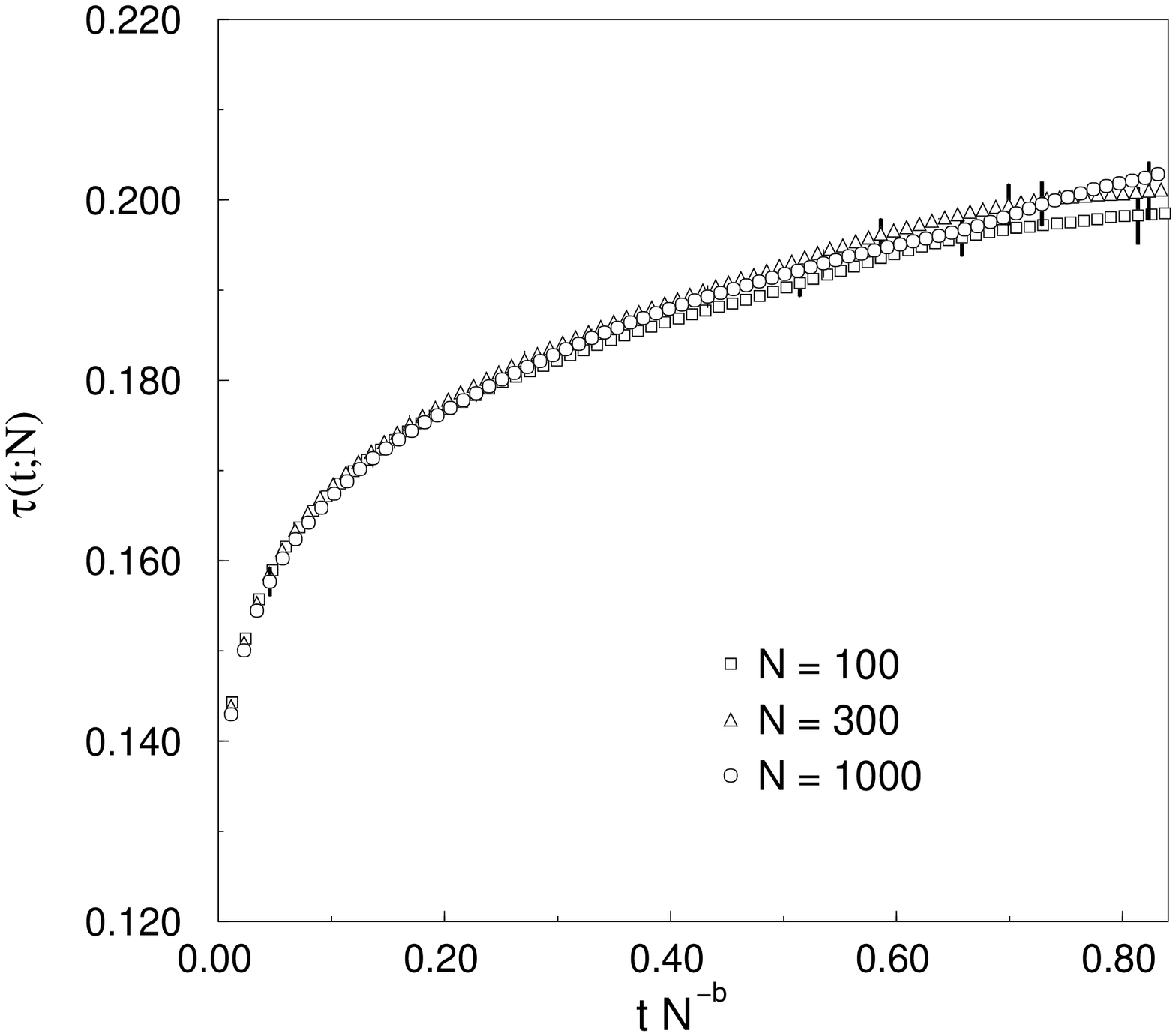,angle=-90,width=0.50\linewidth}
\epsfig{figure=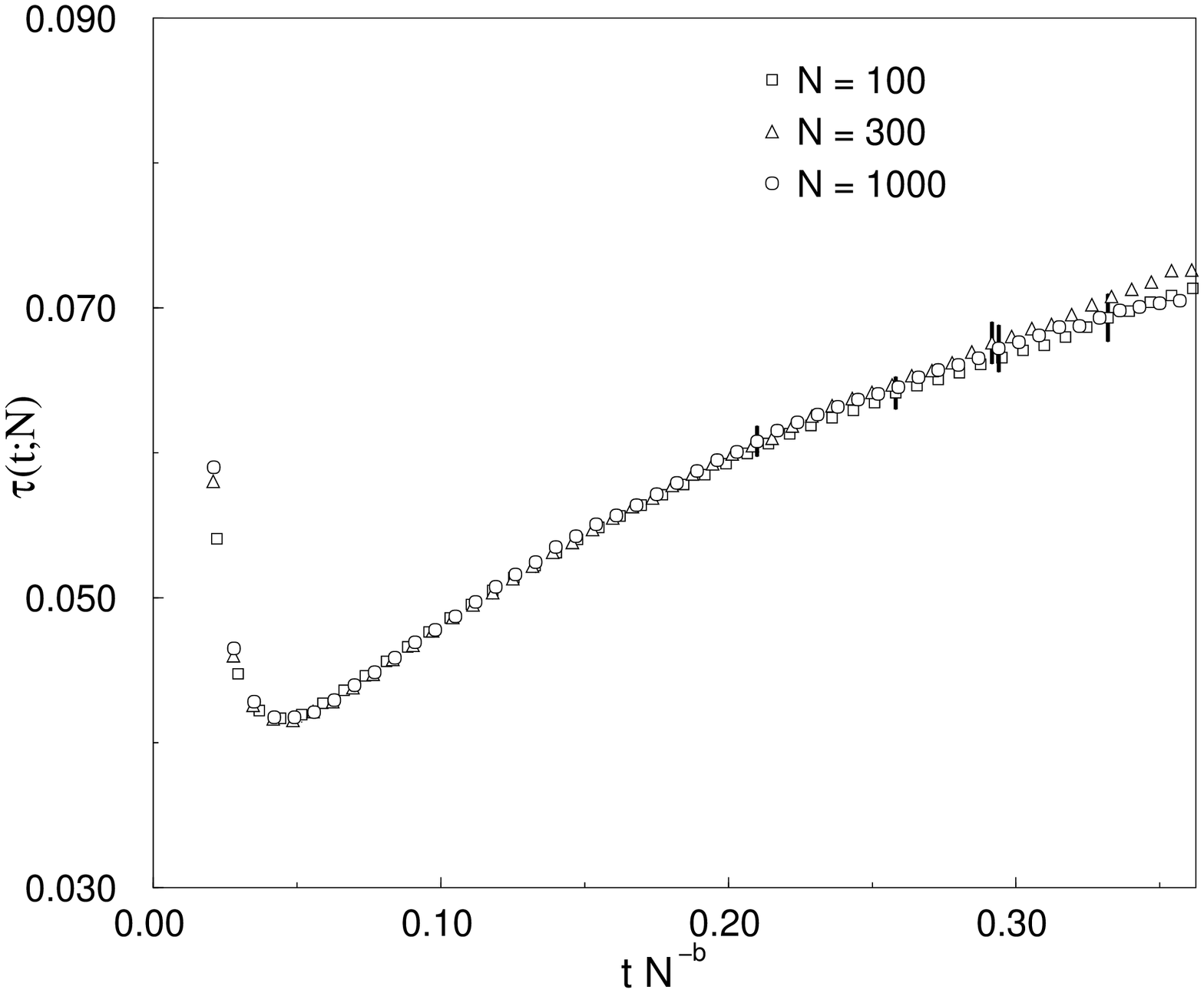,angle=-90,width=0.53\linewidth}
\caption{Dynamic scaling analysis for the EER algorithm in three dimensions: 
plots of 
$\tau_{{\rm scal},A}(t;N)$ vs. $tN^{-b}$. Left frame: 
$A = R^2_g$, $a = 0.0012$, $b = 2.085$; 
right frame: $A = \mathcal E$, $a = 0.210$, $b = 2.03$.}
\label{dyn-scaling-rept-3d}
\end{figure}

In three dimensions, all observables show a very good scaling behavior,
as it can be seen from Fig. \ref{dyn-scaling-rept-3d}:
In all cases the results for the three values of $N$ fall onto a single curve.

{}From the results of Table \ref{rept_scaling_radii} we can compute the 
exponents $z_{\rm exp}$ and $z_{\rm int}$, cf. 
Eqs. \reff{scale_pred1} and \reff{scale_pred2}, and compare them
with the previous results. In three dimensions, the estimates of 
$z_{\rm exp}$ obtained from the scaling analysis 
are in perfect agreement with those of Table \ref{rept_tauexp_fit}. 
In two dimensions instead, only $z_{\rm exp, \cal E}$ is compatible with 
the results of Table \ref{rept_tauexp_fit}.
The estimates of $z_{\rm exp}$ for the radii 
obtained from the 
scaling analysis are significantly lower than those 
obtained from fitting the autocorrelation times.  The origin of the 
discrepancy can be 
understood from Fig. \ref{dyn-scaling-rept-2d-bis}. The exponential time 
is determined by the large-$t$ behavior of $\rho_{AA}(t)$ and in practice 
by the behavior 
in the region in which $t\approx 3 \tau_{{\rm exp},R^2}$, see 
Fig. \ref{effmass-reptation}, which corresponds 
approximately to $t N^{-b} \approx 0.44$. But in this region there is 
no scaling, and thus the corresponding $\tau_{{\rm exp},R^2}$ 
do not scale as $N^{z_{{\rm exp},R^2}}$. For this reason, we believe 
the estimates of Table \ref{rept_tauexp_fit} to be grossly in error. 
Note also that
the estimate we obtain in the scaling analysis,
$z_{{\rm exp},R^2} \approx 2.21$,
is compatible with 
$z_{{\rm exp},R^2} \approx z_{{\rm int},R^2}$, 
a relation that is expected to be true, since the radii are strongly coupled
to the slowest modes of the dynamics.

Then, we can compute $z_{{\rm int},A}$. For the radii, we always have
$a\approx 0$, so that $z_{{\rm int},R^2} \approx z_{{\rm exp},R^2}$
as expected. This confirms the results of Table \ref{rept_int_fit}. 
For the energy, we have instead $a \not = 0$. 
Using Eq. \reff{scale_pred2}, we obtain 
\bea
z_{{\rm int},\mathcal E}  = 1.53 \pm 0.06 && \qquad \qquad d = 2, \\
z_{{\rm int},\mathcal E}  = 1.60 \pm 0.05 && \qquad \qquad d = 3.
\eea
These results are in reasonable agreement with those reported in 
Table \ref{rept_int_fit} if one takes into account that, as we discussed,
the error on those results is of order $0.04$.

In conclusion, putting the results of the different analyses together, 
we obtain in two dimensions:
\bea
&& z_{{\rm exp},R^2} = z_{{\rm int},R^2} = 2.2 \pm 0.1  ,
\nonumber \\
&& z_{{\rm exp},\mathcal E} = 2.15\pm 0.05, \nonumber \\
&& z_{{\rm int},\mathcal E} = 1.60\pm 0.10. 
\eea
In three dimensions we have:
\bea
&& z_{{\rm exp},R^2} = z_{{\rm int},R^2} = 2.07 \pm 0.05  ,
\nonumber \\
&& z_{{\rm exp},\mathcal E} = 2.05\pm 0.05, \nonumber \\
&& z_{{\rm int},\mathcal E} = 1.65\pm 0.05.
\eea
The errors are such to include the results of all analyses.

Note that $z_{{\rm int},\mathcal E} < z_{{\rm exp},\mathcal E}$ 
for the EER dynamics. This may be understood as follows.
The energy fluctuations are essentially due to two causes.
First, there are fluctuations due to local changes of the walk. 
These fluctuations are fast since are due to local and bilocal moves.
Then, there are fluctuations due to changes of the global structure
of the walk. Indeed, there are contributions to the energy
that are due to groups of monomers that are far apart along the walk
but that are near in position space. 
Such contributions vary slowly, 
typically as $\tau_{{\rm exp},R^2}$, and are the origin of the fact
that $z_{{\rm exp},\mathcal E}\approx z_{{\rm exp},R^2}$. 
However, these contributions are very small, and thus give rise to tiny
fluctuations that are negligible 
when considering integrated quantities. Therefore,
$z_{{\rm int},\mathcal E} < z_{{\rm exp},\mathcal E}$. 

All results we have discussed up to now refer to simulations with 
$p=0.5$ and with the first version of the reptation move. 
These choices are of no relevance for the critical exponents, 
but they have of course a strong influence on the amplitudes. 
We have thus tried to see the changes in the dynamics due to a variation 
of $p$ and to a change of the reptation move. To check the role of $p$, 
we have performed simulations with $p=0.1$ and $p=0.9$ with walks of 
length $N=100$ in two dimensions. We find: for $p=0.9$,
$\tau_{{\rm int},R^2_e} = 2184 \pm 32$ and 
$\tau_{{\rm int},\mathcal E} = 1200 \pm 13$;
for $p=0.1$, 
$\tau_{{\rm int},R^2_e} = 8740 \pm 260$ and 
$\tau_{{\rm int},\mathcal E} = 1181 \pm 13$.
This should be compared with the 
results of Table  \ref{rept_tauint}, 
$\tau_{{\rm int},R^2_e} \approx 3110$ and 
$\tau_{{\rm int},\mathcal E} \approx 862$. 
Thus, by increasing $p$, there is a significant speed up of the dynamics of the 
radii---this should be expected, since reptation moves are essentially 
the only ones that change the position 
of the endpoint and are thus those that control the slowest modes of the 
dynamics. Thus, for noninteracting SAWs, for which the energy is not 
an interesting observable, $p$ close to one---but not $p=1$, otherwise
ergodicity is lost---is a good choice for a fast dynamics. 
On the other hand, the dynamics of the energy becomes slower  
both for $p=0.1$ and for $p=0.9$. The fact that $\tau_{{\rm int},\cal E}$
is larger for $p=0.9$ is easy to understand. Indeed, by increasing $p$ 
we decrease the probability of performing L0 and B22 moves that should 
be the most important ones for the energy. 
However, it is clear that also reptation moves are relevant for the energy,
since for $p=0.1$ $\tau_{{\rm int},\cal E}$ is also larger. 
Apparently, for small $p$ the relevant quantity for the 
dynamics of the energy is the number of successful moves, irrespective 
of the type.
Indeed, using the results of App. \ref{App.A.1} and \ref{App.A.2} 
we find $\tau_{{\rm int},\cal E} \approx$ 600 successful iterations
both for $p=0.5$ and $p=0.1$. 

We have also tested the second version of the reptation algorithm,
see App.~\ref{App.A.2}. For random walks, this implementation 
gives $\tau \sim N$ compared to $\tau \sim N^2$ of the first version. 
For the SAW, the two versions are expected to have the same critical 
exponents, but the second one should be faster. We have performed 
a simulation with $p=0.5$ in two dimensions, finding 
$\tau_{{\rm int},R^2_e} = 510\pm 4$ and 
$\tau_{{\rm int},\mathcal E} = 299.8 \pm 1.6$.
These estimates are sensibly smaller than those reported in 
Table \ref{rept_tauint}. For $R^2_e$ the dynamics is faster by a 
factor of 6, and for $\cal E$ by a factor of 3. Clearly, the second 
implementation is the most efficient one and all our simulations should have 
used it. Unfortunately, we thought of this second version only when all 
simulations were completed.

\section{The KER dynamics in two dimensions}  \label{sec6}

The second dynamics we have analyzed is the KER algorithm 
in which end-end reptation moves BEE are replaced by kink-end reptation
moves BKE. It turns out that this dynamics is much slower than the EER one,
and thus we have limited our analysis to shorter walks,
$N=100,300,700$, to two dimensions, and to noninteracting SAWs, 
i.e. $\beta = 0$. We set $p=0.5$.
Again, we measured three radii and the energy. 
The static results agree with those obtained by using the EER dynamics
and discussed in the preceding section and with the 
results of Ref.~\cite{limadsok}.

\begin{table} 
\protect\footnotesize 
\centering 
\begin{tabular}{ccr@{$\,\pm\,$}lr@{$\,\pm\,$}lr@{$\,\pm\,$}l} \hline 
\hline 
 $N$ & iter &
 \multicolumn{2}{c}{\large{$\tau_{{\rm exp},R_g^2}$}} & 
 \multicolumn{2}{c}{\large{$\tau_{{\rm exp},R_e^2}$}} & 
 \multicolumn{2}{c}{\large{$\tau_{{\rm exp},R_m^2}$}} \\ 
\hline 
100 & $2.4\cdot 10^{11}$ & 37600 & 160 & 38000 & 160 & 38620 & 100 \\ 
300 & $8.8\cdot 10^{11}$ & 918000 & 8000 & 958000 & 8000 & 972000 & 8000 \\ 
700 & $1.9\cdot 10^{12}$ & 
   11475000 & 315000 & 11610000 & 315000 & 11250000 & 225000 \\ 
\hline 
\hline 
\end{tabular}
\caption{Exponential autocorrelation times for the KER algorithm
in two and three dimensions. ``iter" is the number of iterations.}
\label{KER_tauexp}
\end{table}

\begin{figure}[!b] 
\hspace{-0.6cm} 
\epsfig{figure=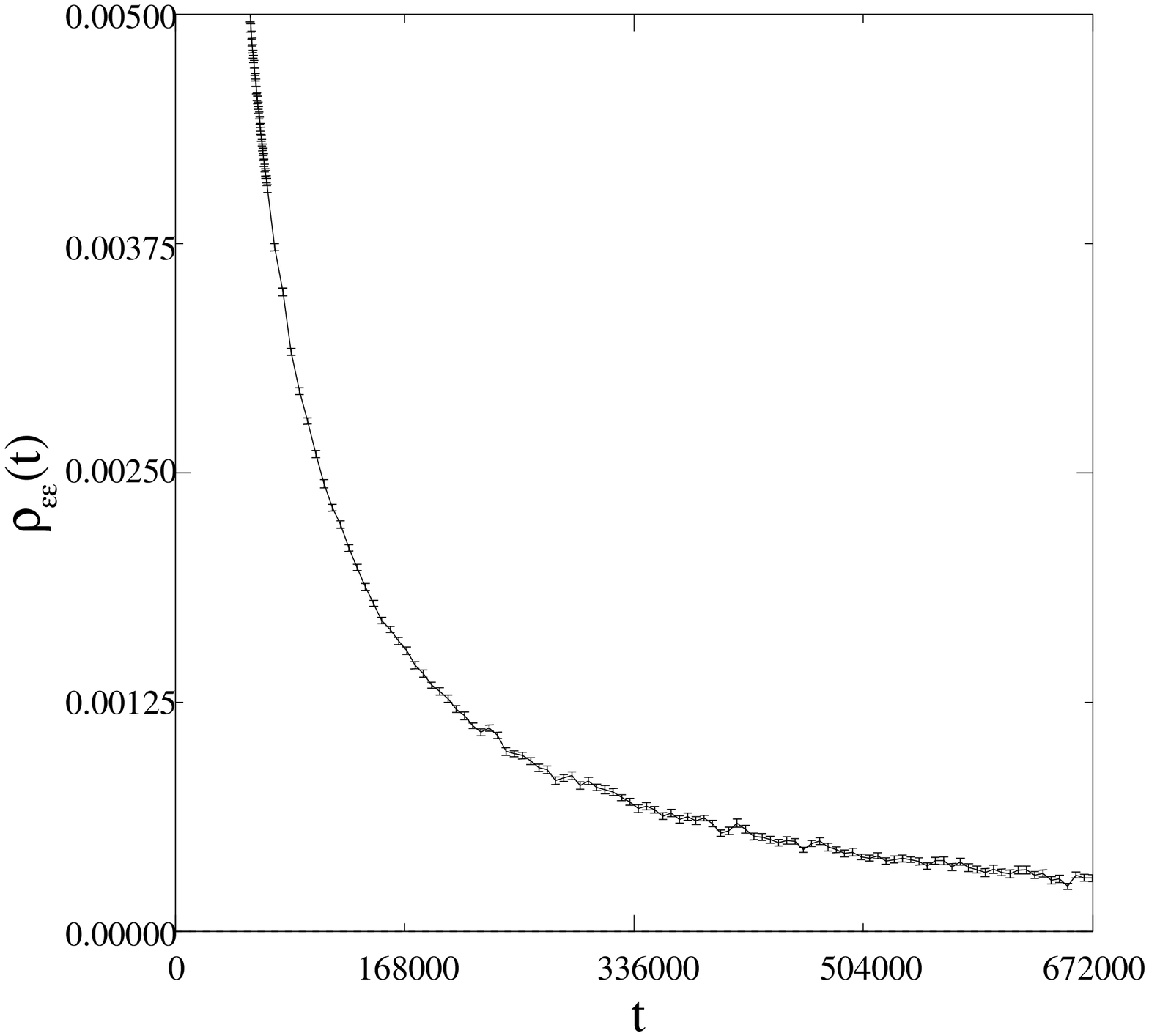,angle=0,width=0.53\linewidth}
\epsfig{figure=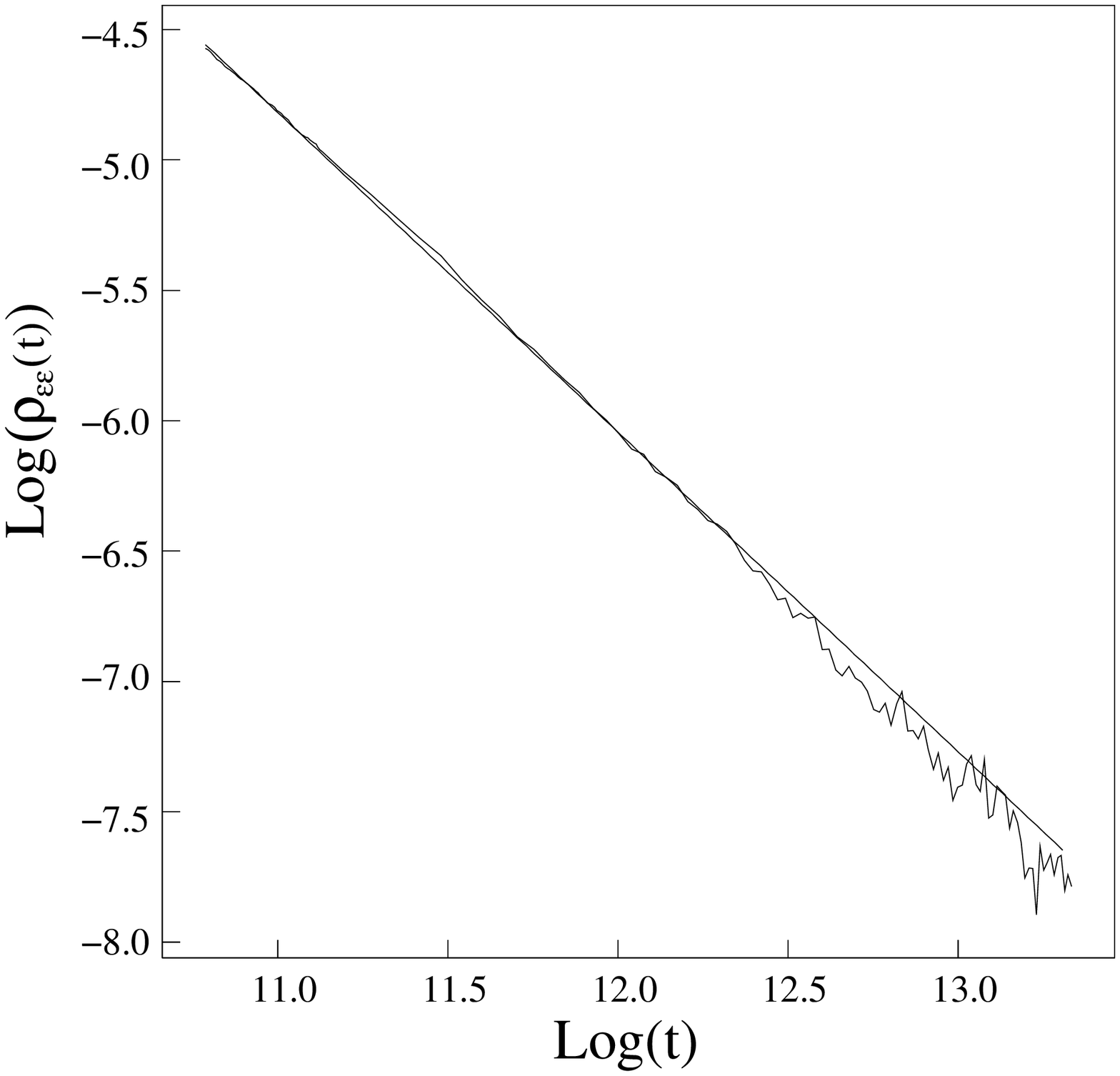,angle=0,width=0.50\linewidth}
\caption{Autocorrelation function $\hat{\rho}_{\cal EE}(t)$ for the 
KER algorithm in two dimensions. Here $N=300$. The straight line 
in the right frame corresponds to $B t^{-\alpha}$, 
$B = 5100$, $\alpha = 1.29$. }
\label{autocorr-fun-KER} 
\end{figure}

As in the preceding section, we have first determined the exponential
autocorrelation times by studying the large-time behavior of the 
effective exponents $\hat{\tau}_{{\rm exp},A}(t;s)$, see Sec. \ref{sec4}.
The results for the radii are reported in Table \ref{KER_tauexp}. 
We have not been able to determine the 
exponential autocorrelation time for the energy $\cal E$. Indeed, 
for the values of $t$ for which $\rho_{\cal EE}(t)$
 is not zero within statistical errors,
$\rho_{\cal EE}(t)$ has a power-law behavior, i.e.
$\rho_{\cal EE}(t)\sim t^{-\alpha}$, with $\alpha\approx 1$-1.3.
This can clearly be seen from Fig. \ref{autocorr-fun-KER} 
where we report $\rho_{\cal EE}(t)$ for $N=300$. 

\begin{table}[!t]
\protect\footnotesize
\centering
\begin{tabular}{cr@{$\,\pm\,$}lr@{$\,\pm\,$}lr@{$\,\pm\,$}l}
\hline
\hline 
& \multicolumn{2}{c}{\large{$\tau_{{\rm exp},R_g^2}$}} &
\multicolumn{2}{c}{\large{$\tau_{{\rm exp},R_e^2}$}} &
\multicolumn{2}{c}{\large{$\tau_{{\rm exp},R_m^2}$}} \\
\hline
$z_{\rm exp}$ & 2.92 & 0.01 & 2.94 & 0.01 & 2.93 & 0.01 \\ 
$B$ & 0.054 & 0.002 & 0.050 & 0.002 & 0.054 & 0.002 \\ 
$\chi^{2}$ & \multicolumn{2}{c}{3.77} & \multicolumn{2}{c}{0.034}   
& \multicolumn{2}{c}{2.46} \\  
\hline 
\hline
\end{tabular}
\caption{Dynamic exponent $z_{\rm exp}$ for the KER algorithm in two and
three dimensions, obtained by fitting $\tau_{\rm exp} = B N^{z_{\rm exp}}$.
The number of degrees of freedom of the fit is 1.}
\label{KER_tauexp_fit}
\end{table}
 
The results for the exponential autocorrelation times are fitted with the 
Ansatz \reff{Ansatzexp}, obtaining the results reported in 
Table~\ref{KER_tauexp_fit}. As before, we cannot perform 
a systematic analysis of the scaling corrections. However, it is 
important to notice that the estimates for the three radii agree within error
bars. This confirms the correctness of the quoted error bars and 
gives the final estimate
\be
z_{{\rm exp},R^2} = 2.93 \pm 0.02.
\label{zexp-KER}
\ee
This estimate is significantly higher than that for the EER dynamics. 
The origin of such a large difference is unclear, since it is difficult 
to see any difference between BEE and BKE moves. Naively, one would have 
expected a BKE move to be equivalent to two BEE moves together with a B22 move.
Since all moves have a finite probability of success as $N\to \infty$,
see Appendix, one would have expected a difference by a constant factor, 
and thus the same critical exponents. 
Such a naive expectation is not true, 
since the exponents are clearly different. 
 
We have then determined the integrated autocorrelation times. 
For the radii we have used the self-consistent windowing method 
of Sec. \ref{sec4}, using $c=15$. The results are reported in 
Table \ref{KER_tauint}. Since $\tau_{{\rm exp},R^2}\sim 3\tau_{{\rm int},R^2}$,
the choice $c=15$ should be enough to avoid systematic errors. 
Instead, the autocorrelation function of the energy decreases 
rapidly and it has a very long tail. In this case, we have chosen a 
much larger valus of $c$, $c=200$, obtaining the results 
that are reported in Table \ref{KER_tauint} as $\tau_{{\rm int},\cal E}$(n.t.).
In spite of $c$ being such a large number, the cutoff value $M$,
cf. Eq.~\reff{tauint-automatic-win},
is still well within the region in which the function decays 
as a power law. For instance, for $N=300$, $M = 670000$ ($\ln M\approx 13.4$)
and $\rho(M) \approx 3.4\cdot 10^{-4}$, see Fig. \ref{autocorr-fun-KER}. 
This should be 
expected since the cutoff $M$ satisfies 
$M\lesssim \tau_{{\rm exp},R^2}\approx 10^6$.
More precisely, $M \approx 3.3\, \tau_{{\rm exp},R^2}$, 
$M \approx 0.7\,\tau_{{\rm exp},R^2}$, and 
$M \approx 0.14\,\tau_{{\rm exp},R^2}$ approximately for 
$N=100,300,700$. Therefore, we expect a sizable contribution from the 
tail of the autocorrelation function, at least for $N=300,700$. 
To take it into account, we use Eq. \reff{tauint-Lietal}, where for 
$\tau_{{\rm exp},\mathcal E}$ we take the average $\tau_{{\rm exp},R^2}$
of the radii. For $N=300$ the correction is of 2.5\%
(2.4\% if we approximate $\rho_{\cal EE}(t) \approx B t^{-1.3}$ as 
obtained from the fit) and for $N=700$ of 20.3\% 
(18.1\% if we approximate $\rho_{\cal EE}(t) \approx B t^{-1.1}$ as
obtained from the fit). The results are reported in 
Table \ref{KER_tauint} as $\tau_{{\rm int},\cal E}$(w.t.). 
The reported error is obtained by summing 
20\% of the contribution of the tail to the original error.
This is an ``ad hoc" prescription which can be shown to work reasonably
in the exactly soluble case of the pivot algorithm for the random walk.
 
\begin{table}[!t] 
\protect\footnotesize 
\centering 
\begin{tabular}{cr@{$\,\pm\,$}lr@{$\,\pm\,$}l
                    r@{$\,\pm\,$}lr@{$\,\pm\,$}l
                                   r@{$\,\pm\,$}l} 
\hline \hline
$N$ & \multicolumn{2}{c}{\large{$\tau_{{\rm int},R_g^2}$}} & 
      \multicolumn{2}{c}{\large{$\tau_{{\rm int},R_e^2}$}} & 
      \multicolumn{2}{c}{\large{$\tau_{{\rm int},R_m^2}$}} & 
      \multicolumn{2}{c}{{\large $\tau_{{\rm int},{\cal E}}$} (n.t.)} & 
      \multicolumn{2}{c}{{\large ${\tau}_{{\rm int},{\cal E}}$} (w.t.)} 
\\ 
\hline 
100 & 15632 & 30 & 20974 & 48 & 27528 & 72 & 847.5 & 0.4 & 847.5 & 0.4\\ 
300 & 341030 & 1640 & 462700 & 2600 & 648500 & 4300  & 3402.6 & 1.6 & 
      3484 & 4 \\ 
700 & 3906500 & 43800 &  5325000 & 70000 & 7942000 & 127000 & 
      9269.0 & 5.1 & 10948 & 90 \\ 
\hline 
$z_{\rm int}$ & 
  2.820 & 0.005 & 2.830 & 0.004 & 2.890 & 0.006 & 1.2330 & 0.0004 &1.289 & 0.0011\\ 
$B$ & 
  0.0360 & 0.0007 & 0.0464 & 0.0010 & 0.0458 & 0.0012 & 2.920 & 0.006
                                   &0.808 & 0.005\\ 
$\chi^{2}$ & \multicolumn{2}{c}{18.5} & 
   \multicolumn{2}{c}{11.8} & \multicolumn{2}{c}{11.8} &
   \multicolumn{2}{c}{4197} & \multicolumn{2}{c}{41.7}\\ 
\hline 
\hline 
\end{tabular}  
\caption{KER algorithm in two dimensions:
integrated autocorrelation times and 
dynamic critical exponents $z_{\rm int}$,
obtained by fitting $\tau_{\rm int} = B N^{z_{\rm int}}$.
The number of degrees of freedom of the fit is 1. 
For the energy $\cal E$, we report two estimates: 
(n.t.) is obtained by using the self-consistent windowing method with 
$c=200$, while (w.t.) is the result obtained including the 
tail contribution \`a la Li {\em et al.} (Ref. \cite{limadsok}).}
\label{KER_tauint}
\end{table} 
 
The results for the autocorrelation times have been fitted 
with the Ansatz \reff{Ansatzint} in order 
to obtain $z_{\rm int}$. For the radii, the quality of the fits is quite 
poor, with a $\chi^2$ of approximately 10-20. Moreover, the 
estimates do not agree within error bars. There are therefore
corrections to scaling larger than the very tiny statistical errors.
By requiring $z_{{\rm int},R^2}$ to coincide for these three quantities,
we obtain finally 
\be
z_{{\rm int},R^2} = 2.85 \pm 0.06,
\ee
that includes all estimates and is compatible with the 
expectation $z_{{\rm int},R^2} = z_{{\rm exp},R^2}$. 

Fits of $\tau_{{\rm int},\cal E}$ are characterized by a very large 
$\chi^2$ and give $z_{{\rm int},\cal E} \approx$ 1.2--1.3, much 
smaller than $z_{{\rm int},R^2}$. As in the EER algorithm, the 
dynamics of the energy is much faster than that of the radii. 
Note that $z_{{\rm int},\cal E}$ is also significantly lower 
that the corresponding exponent for the EER algorithm. 
Again, it is quite difficult to understand intuitively why this happens.
 
\begin{table} 
\protect\footnotesize 
\centering 
\begin{tabular}{cr@{$\,\pm\,$}lr@{$\,\pm\,$}lr@{$\,\pm\,$}lr@{$\,\pm\,$}l} 
\hline \hline
& \multicolumn{2}{c}{$R_g^2$} & \multicolumn{2}{c}{$R_e^2$} & 
 \multicolumn{2}{c}{$R_m^2$} & \multicolumn{2}{c}{$\mathcal{E}$} \\ 
\hline 
$b$ & 2.89 & 0.03 & 2.92 & 0.02 & 2.92 & 0.02 & [2.93 & 0.04]\\ 
$a$ & 0.015 & 0.015 & 0.020 & 0.015 & 0.015 & 0.010 & 0.670 & 0.015\\ 
\hline 
\hline 
\end{tabular} 
\caption{Dynamic exponents $a$ and $b$ for the KER algorithm in two dimensions.
For the energy, we have fixed $b = 2.93\pm 0.04$ and determined the 
corresponding $a$. }
\label{KER_scaling} 
\end{table} 
 
The results reported above are confirmed by a scaling analysis 
using Eq. \reff{scaling_law1}. The results for the radii are reported 
in Table \ref{KER_scaling}. The exponent $a$ is compatible with zero
and $z_{{\rm exp},R^2} = b = 2.91 \pm 0.03$, in agreement 
with the estimate \reff{zexp-KER}. 

Since $\rho_{\cal EE}(t) \sim B t^{-\alpha}$ for the values of $t$ 
we can investigate, we cannot determine $a$ and $b$ independently. 
Thus, we have assumed $z_{{\rm exp},\cal E} = z_{{\rm exp},R^2}$
and then used $b = 2.93\pm 0.04$, cf. Eq. \reff{zexp-KER}---to be 
conservative, we have doubled the error.
Correspondingly, we obtain 
$a = 0.670 \pm 0.015$ and $z_{{\rm int},\cal E} = 0.97 \pm 0.05$,
which is somewhat lower than the estimates of Table \ref{KER_tauint}.
One may think that this is due to our assumption
$z_{{\rm exp},\cal E} = z_{{\rm exp},R^2}$, while there is some evidence 
from the analysis of the EER dynamics
that $z_{{\rm exp},\cal E} < z_{{\rm exp},R^2}$. However, this does not 
explain the difference, since if $z_{{\rm exp},\cal E}$ decreases, also
$z_{{\rm int},\cal E} $ decreases. In order to obtain 
$z_{{\rm int},\cal E} = 1.3$, one should take 
$z_{{\rm exp},\cal E} = b = 3.35$, which is much too large. 
Therefore, the difference should be taken as an indication of the scaling 
corrections.

In conclusion, the KER dynamics has a different critical behavior with 
respect to the EER dynamics. For the critical exponents we have
\bea
&& z_{{\rm exp} ,R^2}  = z_{{\rm int} ,R^2} = 2.90 \pm 0.05 ,\\
&& z_{{\rm int} ,\cal E}  = 1.0 \pm 0.3.
\eea

\section{The EER dynamics at the $\theta$ point in two dimensions} \label{sec7}

Bilocal algorithms are of interest for applications in 
constrained geometries and in the presence of strong interactions
where nonlocal algorithms are inefficient. 

In this section we study the dynamic behavior of the EER algorithm
at the $\theta$ point in two dimensions, by setting 
$\beta = \beta_\theta = 0.665$---we have used the estimate of 
Ref. \cite{GH-95}, see Table \ref{beta-theta}. Here, we have studied 
longer walks than before, $N=100,800,1600,3200$, with large statistics,
see Table \ref{theta_tauexp}. 

\begin{table}
\protect\footnotesize
\centering
\begin{tabular}{ccr@{$\,\pm\,$}lr@{$\,\pm\,$}lr@{$\,\pm\,$}lr@{$\,\pm\,$}l}
\hline\hline
$N$ & iter & \multicolumn{2}{c}{\large{$\tau_{{\rm exp},R_g^2}$}} &
\multicolumn{2}{c}{\large{$\tau_{{\rm exp},R_e^2}$}} &
\multicolumn{2}{c}{\large{$\tau_{{\rm exp},R_m^2}$}} &
\multicolumn{2}{c}{\large{$\tau_{{\rm exp},\cal E}$}} \\                        
\hline
\hline
100 & $5.84 \cdot 10^{10}$ & 5120 & 70 & 5220 & 60 & 5170 & 40 & 4856 & 44\\
800 & $6.28 \cdot 10^{11}$ 
    & 609600 & 15000 & 606000 & 10000 & 620000 & 8000 & 594200 & 9000\\
1600 &$1.81 \cdot 10^{12}$
    & 3165000 & 22000 & 2840000 & 20000 & 3020000 & 20000 & 2936000 &
24000\\
3200 &$1.12 \cdot 10^{13}$
    & 15838000 & 76000 & 13880000 & 50000 & 14990000 & 50000 & 14656000 &
50000\\
\hline
\end{tabular}
\caption{Exponential autocorrelation times for the EER algorithm
in two dimensions at the $\theta$ point. ``iter" is the number of iterations.}
\label{theta_tauexp}    
\end{table}

We have performed the same analyses we have presented in the preceding sections.
First, we have determined the exponential autocorrelation times. 
For all observables, the effective exponent $\hat{\tau}_{{\rm exp},A} (t;s)$
becomes independent of $t$ for $t \approx 2$-$2.5\, {\tau}_{{\rm exp},A}$,
allowing a reliable estimate of the exponential autocorrelation times.
The results are reported in Table \ref{theta_tauexp}. 

Then, we determined the exponent $z_{{\rm exp},A}$ by fitting the 
exponential autocorrelation times to the Ansatz \reff{Ansatzexp}.
The results are reported in Table \ref{theta_tauexp_fit}. 
Clearly, the statistical errors are too small.
Indeed, we expect $z_{{\rm exp},A}$ to be the same 
for all observables---including the energy, that should be strongly
coupled to the slowest modes at the $\theta$ point---and this 
does not happen with the quoted error bars.
By direct comparison of all estimates, we obtain 
the more conservative result
\be
z_{{\rm exp},\cal E} = z_{{\rm exp},R^2} = 2.30 \pm 0.03,
\label{zexp-theta}
\ee
where the error is such to include all estimates.

\begin{table}[!t]
\protect\footnotesize
\centering
\begin{tabular}{cr@{$\,\pm\,$}lr@{$\,\pm\,$}lr@{$\,\pm\,$}lr@{$\,\pm\,$}l}
\hline\hline
  & \multicolumn{2}{c}{\large{$\tau_{{\rm exp},R_g^2}$}} &
\multicolumn{2}{c}{\large{$\tau_{{\rm exp},R_e^2}$}} &
\multicolumn{2}{c}{\large{$\tau_{{\rm exp},R_m^2}$}} &
\multicolumn{2}{c}{\large{$\tau_{{\rm exp},\cal E}$}} \\
\hline
$z_{\rm exp}$ & 2.318 & 0.004 & 2.275 &
0.003 & 2.300 & 0.004 & 2.313 & 0.004 \\
$B$ & 0.126 & 0.006 & 0.148 &
0.004 & 0.128 & 0.005 & 0.116 & 0.006 \\
$\chi^{2}$ & \multicolumn{2}{c}{2.01}  &
\multicolumn{2}{c}{3.64} & \multicolumn{2}{c}{1.24} &
\multicolumn{2}{c}{0.78}\\
\hline
\hline                                                                          
\end{tabular}
\caption{Dynamic exponent $z_{\rm exp}$ for the EER algorithm in two
dimensions at the $\theta$ point, 
obtained by fitting $\tau_{\rm exp} = B N^{z_{\rm exp}}$.
The number of degrees of freedom of the fit is 2.}
\label{theta_tauexp_fit}
\end{table}                                                                     

\begin{table}[!t]
\protect\footnotesize
\centering
\begin{tabular}{cr@{$\,\pm\,$}lr@{$\,\pm\,$}lr@{$\,\pm\,$}lr@{$\,\pm\,$}l} 
\hline\hline
$N$ & \multicolumn{2}{c}{\large{$\tau_{{\rm int},R_g^2}$}} & 
      \multicolumn{2}{c}{\large{$\tau_{{\rm int},R_e^2}$}} & 
      \multicolumn{2}{c}{\large{$\tau_{{\rm int},R_m^2}$}} & 
      \multicolumn{2}{c}{\large{$\tau_{{\rm int},\mathcal E}$}} \\
\hline
100 & 4520 & 10  &  3166.2 & 5.8 & 3082.8 & 5.4 & 2504.4 & 4.0\\
800 & 559500 & 4100 &  362300 & 2100 &  357900 & 2100 & 222800 & 1000\\
1600 & 2890400 & 28300 &  1779500 & 13700 & 1778000 & 13600 & 1029500 &
6000\\
3200 & 13953000 & 120000 & 8630000 & 59000 & 8698000 & 59000 & 4185000 &
20000\\ 
\hline
$z_{\rm int}$ & 2.321 & 0.002 & 2.282 & 0.002 &
2.291 & 0.002 & 2.1510 & 0.0011\\
$B$ & 0.1020 & 0.0013 & 0.0864 & 0.0008 &
0.0800 & 0.0010 & 0.1260 & 0.0008\\
$\chi^{2}$ & \multicolumn{2}{c}{9.0}  & 
             \multicolumn{2}{c}{1.22} & 
             \multicolumn{2}{c}{3.73} & 
             \multicolumn{2}{c}{149}\\
\hline
\hline
\end{tabular}
\caption{EER algorithm in two dimensions at the $\theta$ point:
integrated autocorrelation times and
dynamic critical exponents $z_{\rm int}$,
obtained by fitting $\tau_{\rm int} = B N^{z_{\rm int}}$.
The number of degrees of freedom of the fit is 2.  }
\label{theta_tauint}
\end{table}

We have analogously determined the integrated autocorrelation times 
using the self-consistent windowing method with $c=15$. The 
results for $\tau_{{\rm int},A}$ are reported in Table \ref{theta_tauint}. 
Notice that in this case the integrated autocorrelation times
for the energy are close to those of the radii, as it should be expected, 
since at the $\theta$ point also the energy is a ``slow" variable. 
In all cases, $\tau_{{\rm exp},A}\sim$ 1-3$\, \tau_{{\rm int},A}$ and thus 
the choice $c=15$ should give a small systematic error 
due to the truncation of the autocorrelation functions. 

\begin{figure}[!b]
\begin{center}
\epsfig{figure=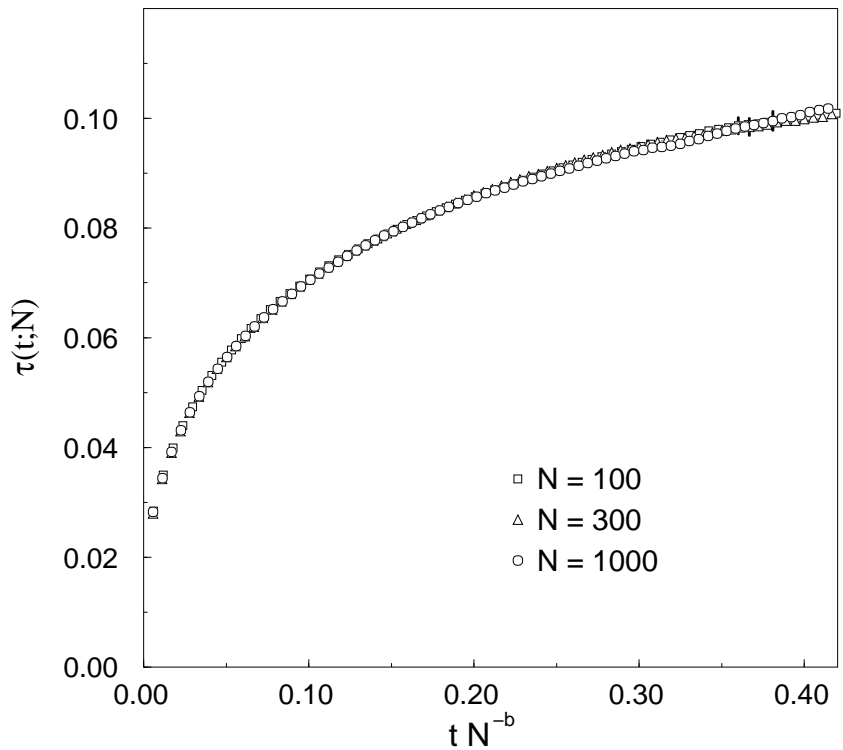,angle=-90,width=0.70\linewidth}
\end{center}
\caption{Dynamic scaling analysis for the EER algorithm in two dimensions
at the $\theta$ point: plots of $\tau_{{\rm scal},R^2_m}(t;N)$ vs. $tN^{-b}$.
Here $a = 0.0015$, $b = 2.31$. }
\label{dyn-scaling-theta}
\end{figure}         

The integrated  autocorrelation times have been fitted with the 
Ansatz $B N^{z_{\rm int}}$, in order to compute $z_{{\rm int},A}$.
The results are reported in Table \ref{theta_tauint}. 
In all cases, the purely statistical errors we have reported 
are too small. For the radii, the exponent $z_{{\rm int}}$ should be 
the same, and thus the error is at least a factor of ten larger.
Comparing the estimates of Table \ref{theta_tauint}, we arrive 
at the final result
\be
z_{{\rm int},R^2} = 2.30 \pm 0.03,
\ee
that, by comparing with Eq. \reff{zexp-theta}, gives
$z_{{\rm int},R^2} = z_{{\rm exp},R^2}$, as expected.

The result for $\cal E$ is somewhat lower, but the very large $\chi^2$ 
indicates that corrections to scaling are significant. In order to
see if there is a systematic trend we have determined an effective 
$z_{{\rm int},\cal E}$, by computing 
\be
\hat{z}_{{\rm int},\cal E}(N_1,N_2) = 
  \left[\ln{\tau_{{\rm int},\cal E}(N_1)\over \tau_{{\rm int},\cal E}(N_2)}
  \right]\, 
  \left(\ln{N_1\over N_2}\right)^{-1}.
\ee
We obtain
\bea
\hat{z}_{{\rm int},\cal E}(100,800) &=& 2.158 \pm 0.003, \\
\hat{z}_{{\rm int},\cal E}(800,1600) &=& 2.208 \pm 0.015 \\
\hat{z}_{{\rm int},\cal E}(1600,3200) &=& 2.023 \pm 0.015 
\eea
It is difficult to observe a systematic trend, but in any case a systematic 
increase towards $2.30$ seems to be excluded. 
On the contrary, the data seem to indicate that $z_{{\rm int},\cal E}$ 
decreases below the value of Table \ref{theta_tauint}.
Thus, also at the $\theta$ point we have 
$z_{{\rm int},\cal E} < z_{{\rm exp},\cal E}$, although the 
difference is much smaller than in the case $\beta = 0$.

These results are confirmed by a scaling analysis. In 
Fig.~\ref{dyn-scaling-theta} we report the scaling variable 
$\tau_{{\rm scal},R^2_m}(t;N)$.
We observe a very good scaling behavior and correspondingly we are able 
to obtain quite reliable estimates of the critical exponents 
$a$ and $b$. The same good behavior is observed for all observables.
The estimates of $a$ and $b$ are reported in Table 
\ref{theta_scaling}. For all observables, $b$ is compatible with 
the estimates of Table \ref{theta_tauexp_fit}, confirming 
the estimate \reff{zexp-theta}. For the radii, $a = 0$, 
in agreement with Table \ref{theta_tauint}. For the energy, 
$a$ is clearly nonvanishing, confirming that 
$z_{{\rm int},\cal E} < z_{{\rm exp},\cal E}$. 
Using Eq. \reff{scale_pred2}, we have 
$z_{{\rm int},\cal E} = 2.15 \pm 0.03$, which is in agreement with
the previous results.

\begin{table}
\protect\footnotesize
\centering
\begin{tabular}{cr@{$\,\pm\,$}lr@{$\,\pm\,$}lr@{$\,\pm\,$}lr@{$\,\pm\,$}l} 
\hline\hline
& \multicolumn{2}{c}{$R_g^2$} & \multicolumn{2}{c}{$R_e^2$} &
 \multicolumn{2}{c}{$R_m^2$} & \multicolumn{2}{c}{$\mathcal{E}$} \\
\hline
$b$ & 2.315 & 0.015 & 2.30 & 0.02 & 2.31 & 0.02 & 2.31 & 0.02\\
$a$ & 0.0015 & 0.0015 & 0.002 & 0.002 & 0.0015 & 0.0015 & 0.07 & 0.01\\
\hline
\hline
\end{tabular}
\caption{Dynamic exponents $a$ and $b$ for the EER algorithm in two 
dimensions at the $\theta$ point.}
\label{theta_scaling}
\end{table}

\section{Conclusions} \label{sec8}

The simulations we have presented show that the reptation dynamics is 
quite successful, even at the $\theta$ point. The values of $z$ we 
have found are only marginally higher than 2, which is the best possible 
behavior for a dynamics that involves local and bilocal moves. 
Of course, for a practical implementation one may want to explore 
several variants that, although do not change the critical behavior,
may still speed up the dynamics by a constant (large) factor. 
First of all, in practical implementations it is important to use the 
second version of the reptation dynamics (see App. \ref{App.A.2}). 
Second, the B22 moves are quite rarely performed and in any case much less 
than the kink-end/end-kink moves. For instance, in two dimensions 
at $\beta = 0$, B22 (resp. BKE) moves are performed with probability 
0.08 (resp. 0.13). Moreover, BKE moves appear to be quite successful 
in speeding up the dynamics of the energy, that is one of the slow
variables in the presence of interactions. Therefore, in the compact regime
an efficient dynamics can be obtained by mixing together: 
(i) the reptation move; (ii) the BKE move; (iii) purely local moves L0 and L1.
A purely local algorithm that leaves the correct measure invariant can be 
obtained from that described in App. \ref{App.A.1} by setting 
$p(0) = 1/2$ in all dimensions and $p(22) = 0$. In this case, it is 
convenient to include also crankshaft moves, see Sec. 4.1 of 
Ref. \cite{Caracciolo-etal_2000}. Such an implementation of the EER 
algorithm should be the method of 
choice for fixed $N$ simulations in the compact regime.


\appendix
\section{The basic moves} \label{App.A}

In this appendix we introduce the basic moves that we use in
our simulation:
\begin{itemize}
\item[(i)] The kink-kink local/bilocal move;
\item[(ii)] The reptation move;
\item[(iii)] The kink-end/end-kink reptation move.
\end{itemize}
In Ref. \cite{Caracciolo-etal_2000} it was shown that 
moves (iii) are enough to obtain an ergodic algorithm. 
In two dimensions one can limit oneself to consider only 
moves (i), but this algorithm is inefficient because of the 
slow motion of the endpoint. Reptation moves are never ergodic 
because of the possibility that the endpoints get trapped. 

\subsection{Kink-kink local/bilocal move} \label{App.A.1}

In this section we define the kink-kink local/bilocal 
move \cite{Caracciolo-etal_2000}. In order to
describe the algorithm it is important to classify the possible 
configurations of three successive links (see Fig. \ref{lowerg}):
\begin{figure}  
\centering
\vspace{0.3cm}
\epsfig{figure=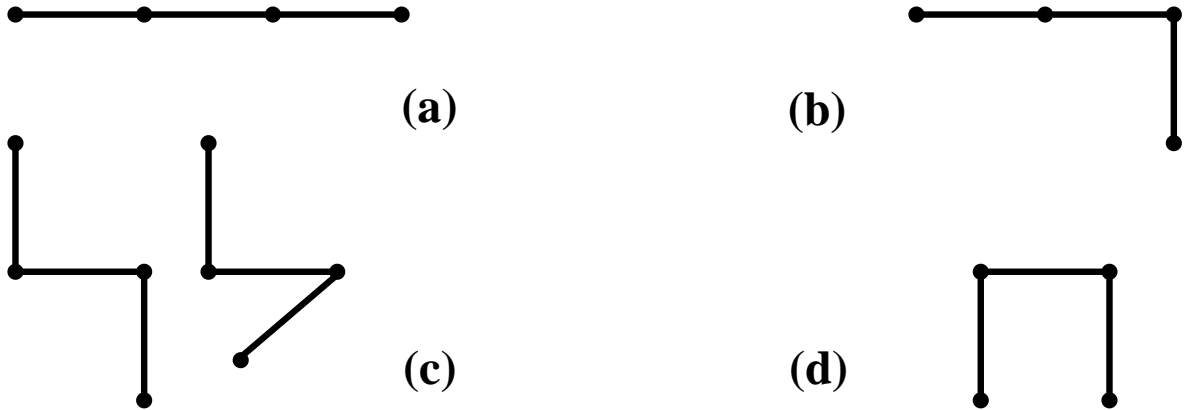,angle=0,width=0.95\linewidth}
\vspace{0.5cm}
\caption{
Configurations of three consecutive links: 
(a) configuration of type  $\textsf{I}$; (b)  configuration of type 
$\textsf{L}$; (c)  configuration of type $\textsf{S}$; 
(d)   configuration of type $\textsf{U}$.}
\label{lowerg}  
\end{figure}  
\begin{enumerate}
\item the bonds have the same direction (\textsf{I}
configuration);
\item two consecutive bonds have the same direction, while the third 
one is perpendicular to them (\textsf{L} configuration);
\item the first and the third bond are perpendicular to the second one, and
they are either parallel or perpendicular to each other (\textsf{S}
configuration);
\item the first and the third bond are perpendicular to the second one, and
they are antiparallel to each other (\textsf{U} configuration).
\end{enumerate}

An iteration works as follows:
\begin{itemize}
\item{Step 1.} Choose a random site $i$ of the current walk $\omega$,
$0\le i \le N$. If $i = N$, propose an L1 move and go to step 5.

\item{Step 2.} Determine the configuration of the subwalk 
$\omega[i-1,i+2]$. If $i = N-1$, we imagine adding a link 
$\Delta\omega(N)$ parallel to $\Delta\omega(N-1)$, so that 
the possible configurations are of type \textsf{L} and 
\textsf{I}. Analogously, if $i = 0$, we imagine adding a 
link $\Delta\omega(-1)$ parallel to $\Delta\omega(0)$.

\item{Step 3.} Draw a random number $r$, uniformly distributed in [0,1].
Depending on the configuration of 
$\omega[i-1,i+2]$,  do the following:
\begin{enumerate}
\item{\textsf{I}}: If $r > (2 d - 2) p(22)$, perform a 
null transition and the iteration ends. Otherwise, go the next step.
\item{\textsf{L}}: If $r > (2 d - 3) p(22) + p(0)$, 
perform a null transition and the iteration ends. 
If $(2 d - 3) p(22) < r < (2 d - 3) p(22) + p(0)$, propose an
L0 move and go to step 5. Otherwise, go to the next step.
\item{\textsf{S}}: If $r > (2 d - 4) p(22) + 2 p(0)$ 
perform a null transition and the iteration ends. 
If  $(2 d - 4) p(22) < r < (2 d - 4) p(22) + 2 p(0)$ propose 
an L0 move: there are two possibilities which are chosen
amongst randomly; then go to step 5.
Otherwise, go to the next step.
\item{\textsf{U}}: Go to the next step.
\end{enumerate}

\item{Step 4.} 
Choose a second integer $j$ uniformly in the disjoint intervals, 
$-1\le j \le N$, $j\not = i-1,i, i+1$.
If $j = -1,N$ make a null transition and the iteration ends. Otherwise, 
depending on the configuration of $\omega[i-1,i+2]$, do the following:
\begin{itemize}
\item $\omega[{i-1,i+2}]$ is of type \textsf{I}, \textsf{S}, \textsf{L}:
if $j=0$ or $j = N - 1$, or if $\omega[j-1,j+2]$ is not of type 
\textsf{U} perform a null transition and the iteration ends. 
Otherwise, propose a B22 
move, cutting the kink $\omega[j-1,j+2]$ and adding it to 
$\omega[i,i+1]$ in one of the possible directions \cite{foot3}.
Then, go to the next step.
\item $\omega[{i-1,i+2}]$ is of type \textsf{U}:
according to the configuration of $\omega[j-1,j+2]$ (if $j=0,N-1$ imagine
adding links as before) do the following:
\begin{enumerate}
\item $\omega[j-1,j+2]$ is of type \textsf{I}: 
If $r < (2 d - 2)p(22)$ (note that the random number $r$ appearing here is 
the same used in Step 3.), propose a B22 move: cut the kink
$\omega[i-1,i+2]$ and add it on top of $\omega[j,j+1]$ 
in a possible random direction, and then go to step 5. 
Otherwise, perform a null transition and the iteration ends.
\item $\omega[j-1,j+2]$ is of type \textsf{L}: 
If $r < (2 d - 3)p(22)$, propose a B22 move: cut the kink
$\omega[i-1,i+2]$ and add it  on top of $\omega[j,j+1]$ 
in a possible random direction, and then go to step 5. 
Otherwise, perform a null transition and the iteration ends.
\item $\omega[j-1,j+2]$ is of type \textsf{S}: 
If $r < (2 d - 4)p(22)$ propose a B22 move: cut the kink
$\omega[i-1,i+2]$ and add it on top of $\omega[j,j+1]$ 
in a possible random direction, and then go to step 5. 
Otherwise, perform a null transition and the iteration ends.
\item $\omega[j-1,j+2]$ is of type \textsf{U}: 
If $r < (2 d - 3)p(22)$, propose a B22 move: cut the kink
$\omega[i-1,i+2]$ and add it on top of $\omega[j,j+1]$ 
in a possible random direction, and then go to step 5; 
if $(2 d - 3)p(22) < r < 2 (2 d - 3)p(22)$, propose a B22 move: cut the kink
$\omega[j-1,j+2]$ and add it on top of $\omega[i,i+1]$ 
in a possible random direction, and then go to step 5. 
Otherwise, perform a null transition and the iteration ends.
\end{enumerate}
\end{itemize}
\item{Step 5.} Check for self-avoidance. If the proposed new walk
is self-avoiding keep it, otherwise perform a null transition.
\item{Step 6.} Compute the difference in energy between the old and the 
new walk and perform a Metropolis test. 
\end{itemize}

The algorithm we have presented depends on the probabilities 
$p(0)$ and $p(22)$, that are the probabilities of 
L0 and B22 moves respectively. As discussed in Ref. \cite{Caracciolo-etal_2000},
the fastest dynamics is obtained by setting
\begin{eqnarray}
p(0) & = & {d-1\over 4 d - 6}, \\
p(22) & = & \frac{1}{4d - 6}.
\end{eqnarray}
In two dimensions $p(0) = p(22) = 1/2$, while in three dimensions
$p(0) = 1/3$ and $p(22) = 1/6$.

\begin{table}[!t]
\protect\footnotesize 
\centering 
\begin{tabular}{ccrcr} \hline \hline
$d$ & $N$ & & \multicolumn{1}{c}{L0, L1} & \multicolumn{1}{c}{B22} \\ 
\hline  
  & 100 & $p_{\rm pr}$ & 0.496 & 0.0952  \\  
  &     & $p_{\rm SA}$ & 0.751 & 0.852   \\  
  & 300 & $p_{\rm pr}$ & 0.499 & 0.0963 \\ 
2 &     & $p_{\rm SA}$ & 0.751 & 0.851  \\  
  & 700 & $p_{\rm pr}$ & 0.500 & 0.0966 \\ 
  &     & $p_{\rm SA}$ & 0.751 & 0.851 \\ 
  &1000 & $p_{\rm pr}$ & 0.500 & 0.0967  \\  
  &     & $p_{\rm SA}$ & 0.751 & 0.851 \\  
\hline  
  & 100 & $p_{\rm pr}$ & 0.443 & 0.0850 \\ 
  &     & $p_{\rm SA}$ & 0.800 & 0.882  \\ 
3 & 300 & $p_{\rm pr}$ & 0.446 & 0.0864 \\ 
  &     & $p_{\rm SA}$ & 0.798 & 0.878   \\ 
  &1000 & $p_{\rm pr}$ & 0.447 & 0.0869  \\ 
  &     & $p_{\rm SA}$ & 0.798 & 0.877 \\ 
\hline  \hline
\end{tabular} 
\caption{Proposal probability $p_{\rm pr}$ and probability 
$p_{\rm SA}$ that the proposed walk is self-avoiding for local 
(L0 and L1) and bilocal (B22) moves. We consider noninteracting 
SAWs in two and three dimensions.}
\label{KKbilocal_accept} 
\end{table}

\begin{table}[!t]
\protect\footnotesize
\centering
\begin{tabular}{crcr} \hline\hline
 \multicolumn{1}{c}{$N$} & & 
 \multicolumn{1}{c}{L0, L1} & \multicolumn{1}{c}{B22} \\
\hline 
     & $p_{\rm pr}$  & 0.463 & 0.158 \\
100  & $p_{\rm SA}$  & 0.429 & 0.525  \\ 
     & $p_{\rm Met}$ & 0.701 & 0.610 \\
\hline
     & $p_{\rm pr}$  & 0.462 & 0.162 \\
800  & $p_{\rm SA}$  & 0.394 & 0.416 \\
     & $p_{\rm Met}$ & 0.711 & 0.561 \\
\hline
     & $p_{\rm pr}$  & 0.462 & 0.163 \\
1600 & $p_{\rm SA}$  & 0.389 & 0.399 \\
     & $p_{\rm Met}$ & 0.714 & 0.556 \\
\hline
     & $p_{\rm pr}$  & 0.462 & 0.163 \\
3200 & $p_{\rm SA}$  & 0.385 & 0.388 \\
     & $p_{\rm Met}$ & 0.716 & 0.553 \\
\hline \hline
\end{tabular}
\caption{Proposal probability $p_{\rm pr}$, probability 
$p_{\rm SA}$ that the proposed walk is self-avoiding,  and 
probability $p_{\rm Met}$ that the proposed self-avoiding walk is 
accepted in the Metropolis test. For local 
(L0 and L1) and bilocal (B22) moves. We consider 
SAWs in two dimensions at the $\theta$ point.}
\label{KKbilocal_theta_accept} 
\end{table}

It is interesting to compute the probability of a successful move. 
For noninteracting SAWs such a probability is the product of two terms: 
the probability $p_{\rm pr}$ that a given move is proposed and the 
probability $p_{\rm SA}$ that the proposed walk is self-avoiding.
At the $\theta$-point one must additionally multiply by the 
probability that the Metropolis test is successful. 
Numerical estimates of these probabilities are reported in 
Tables \ref{KKbilocal_accept} and \ref{KKbilocal_theta_accept}. 
Note that they have a very weak $N$ dependence and clearly approach a constant
value as $N$ goes to infinity.

The probability $p_{\rm pr}$ can be easily computed by using the probability 
of occurrence of the four configurations
$\textsf{I}$, $\textsf{L}$, $\textsf{U}$,
and $\textsf{S}$ defined in Fig.~\protect\ref{lowerg}. 
Indeed, 
\begin{eqnarray}
p_{\rm pr,B22} &=& 2 p(22) p(\textsf{U})
   \left[ (2 d - 2) p(\textsf{I}) +
          (2 d - 3) (p(\textsf{L}) + p(\textsf{U})) +
          (2 d - 4) p(\textsf{S}) \right],
\\
p_{\rm pr,L} &=& p(0) [ p(\textsf{L}) + 2 p(\textsf{S}) ].
\end{eqnarray}                                                                  
Using the numerical results of Table \ref{threelinks-conf}, 
we obtain $p_{\rm pr} = 0.097$, 0.087, 0.164 for B22 moves and 
($d=2$, $\beta = 0$), ($d=3$, $\beta = 0$), and 
($d=3$, $\beta = \beta_\theta)$ respectively. 
For local moves we obtain correspondingly 
$p_{\rm pr} = 0.500$, 0.462, 0.447. These results are in good agreement 
with the numerical ones of 
Tables \ref{KKbilocal_accept} and \ref{KKbilocal_theta_accept}. 

\begin{table}[tpb]
\protect\footnotesize
\centering
\begin{tabular}{lccc} 
\hline\hline
& $\beta=0$, $d=2$ & $\beta=\beta_\theta$, $d=2$ &
  $\beta=0$, $d=3$ \\
\hline
\vphantom{,} &&& \\[-3mm]
\large{\textsf{I}} & 0.151 & 0.128 & 0.051 \\
\large{\textsf{L}} & 0.480 & 0.455 & 0.354 \\
\large{\textsf{U}} & 0.109 & 0.183 & 0.102 \\
\large{\textsf{S}} & 0.260 & 0.234 & 0.493 \\
\hline
\hline
\end{tabular}
\caption{Probabilities for $N\to \infty$ of the occurence of the 
four configurations $\textsf{I}$, $\textsf{L}$, $\textsf{U}$,
and $\textsf{S}$ defined in Fig.~\protect\ref{lowerg}. Results 
for noninteracting SAWs  
in two and three dimensions and for SAWs at the $\theta$ point
in two dimensions.}
\label{threelinks-conf}
\end{table}

Using the above presented results, we can compute the probability 
of a successful move. They are reported in Table 
\ref{KKbilocal_success}. Note that at the $\theta$ point the probability 
of a null transition is quite large and in particular B22 moves 
are quite rarely performed. 

\begin{table}[tpb]
\protect\footnotesize
\centering
\begin{tabular}{lcccc} 
\hline\hline
$d$ &  $\beta$ & L0,L1  & B22  & null \\
\hline
2 & 0            & 0.38 & 0.08 & 0.54 \\
3 & 0            & 0.36 & 0.08 & 0.56 \\
2 &$\beta_\theta$& 0.13 & 0.03 & 0.84 \\
\hline
\hline
\end{tabular}
\caption{Probability of the different moves for different 
$\beta$ and $d$. }
\label{KKbilocal_success}
\end{table}

\subsection{Reptation move} \label{App.A.2}

There are two different implementation of 
the reptation (or slithering-snake) move.
The first one, which satisfies detailed balance,
works as follows [{\em Version 1}]:
\begin{itemize}
\item Step 1. With probability 1/2 delete $\omega[N-1,N]$ and add a
new link at the beginning of the walk; 
otherwise, delete $\omega[0,1]$ and add a new 
link at the end of the walk.

\item Step 2. Check if the new walk is self-avoiding. If it is keep it, 
otherwise perform a null transition.

\item{Step 3.} Compute the difference in energy between the old and the 
new walk and perform a Metropolis test. 
\end{itemize}

A second version uses an additional flag which specifies which 
of $\omega(0)$ and $\omega(N)$ is the ``active" endpoint.
It works as follows [{\em Version 2}]:
\begin{itemize}
\item Step 1. Delete one bond at the ``active" endpoint and append a new 
one at the opposite end of the walk.

\item Step 2. If the new walk is self-avoiding keep it, otherwise 
stay with the old walk, and change the flag, switching the active endpoint.

\item{Step 3.} Compute the difference in energy between the old and the 
new walk and perform a Metropolis test. 
\end{itemize}
This algorithm\cite{foot4} 
does not satisfy detailed balance, but it satisfies the 
stationarity condition generating the correct probability distribution.

\begin{table}[!t]
\protect\footnotesize 
\centering 
\begin{tabular}{ccrr} \hline 
\hline  
$d$ & \multicolumn{1}{c}{$N=100$} & \multicolumn{1}{c}{$N=300$} & 
      \multicolumn{1}{c}{$N=1000$}\\ 
\hline 
2 & 0.882 & 0.880 & 0.880  \\
3 & 0.938 & 0.937 & 0.937  \\
\hline  
\hline
\end{tabular} 
\caption{Probability $p_{\rm SA}$ that the proposed walk is self-avoiding 
for the reptation move in two and three dimensions. We consider
noninteracting SAWs.} 
\label{rept_accept} 
\end{table}

\begin{table}
\protect\footnotesize
\centering
\begin{tabular}{crrrr}
\hline\hline
 & \multicolumn{1}{c}{$N=100$}  &
   \multicolumn{1}{c}{$N=800$}  &
   \multicolumn{1}{c}{$N=1600$}  &
   \multicolumn{1}{c}{$N=3200$}  \\
\hline
$p_{\rm SA}$  & 0.643 & 0.566 & 0.551 & 0.540 \\
$p_{\rm Met}$ & 0.697 & 0.663 & 0.658 & 0.654 \\
\hline \hline
\end{tabular}
\caption{Probability 
$p_{\rm SA}$ that the proposed walk is self-avoiding and 
probability $p_{\rm Met}$ that the proposed self-avoiding walk is accepted in 
the Metropolis test. For reptation moves. We consider
SAWs in two dimensions at the $\theta$ point.}
\label{rept_theta_accept} 
\end{table}

It is interesting to compute the probability of success of a reptation move.
In the absence of interactions it is simply given by the probability 
that the proposed walk is self-avoiding. Such a probability is reported 
in Table \ref{rept_accept}. The reptation move is quite successful, 
being accepted with high probability in both two and three dimensions. 

At the $\theta$ point, we must also consider the probability that the 
proposed walk passes the Metropolis test. Numerical results are 
reported in Table \ref{rept_theta_accept}. Since the walk is more compact, 
$p_{\rm SA}$ is lower than in the noninteracting case, although still 
quite large. Multiplying the two probabilities we see that the 
reptation move is accepted in 35\% of the cases. Note that this 
probability is larger than the probability of a local or bilocal B22 
move, see Table \ref{KKbilocal_success}.

\subsection{Kink-end/end-kink move} \label{App.A.3}

The kink-end/end-kink move uses BKE moves (see Fig.~\ref{kink-end}).
It consists of the following steps:
\begin{itemize}
\item{Step 1.} Choose a random site $i$ of the current walk with 
$0\le i \le N - 2$.
\item{Step 2.} Propose an end-kink move with probability $(2d-2) p$ or a
kink-end move with probability $(2d-1)^2 p$. In the first case
delete the last two bonds of the walk and insert 
a kink on the bond $\Delta\omega(i)$ in one of the $(2 d - 2)$
possible orientations. In the second case,
if $i \not= 0$ and $\omega[i-1,i+2]$ is a kink,
remove it and attach two bonds at the end of the walk
in one of the $(2d - 1)^2$ possible ways. Otherwise, perform a
null transition and the iteration ends.
\item {Step 3.} Check if the proposed walk is self-avoiding. If it is keep it,
otherwise make a null transition.
\item{Step 4.} Compute the difference in energy between the old and the 
new walk and perform a Metropolis test. 
\end{itemize}
The constant $p$ is given by \cite{Caracciolo-etal_2000}
\be
p  =  \frac{1}{(2d - 1)^2 + (2d -2)}.
\ee
We obtain 
$p = 1/11$ in $d=2$, and $p = 1/29$ in $d=3$.
A slightly more efficient implementation is discussed in 
Ref. \cite{Caracciolo-etal_2000}.

We have computed numerically the probability that a kink-end or an end-kink 
move is accepted. We find 0.140, 0.138, 0.138 for $N=100,300,700$ respectively.
The probability of a null transition is therefore quite large, much larger than
for a kink-kink bilocal move. Note however, that a kink-end/end-kink 
move is performed  more often than a B22 move and thus this type of 
moves should be slightly more efficient in updating the part of the 
walk that is far from the endpoints than B22 moves.



\begin{thebibliography}{199}
\bibitem{foot1}
It should be noted that the
ergodicity problem is less severe for reptation than for local
algorithms. For instance, for reptation one can prove
(see Ref. \cite{Madras-Sokal_87} and Proposition 9.2.3 in
Ref. \cite{Madras-Slade_book})
that the fraction of walks belonging to the ergodicity class of the
straight rod is larger than $\sim N^{-(\gamma - 1)/2}$. On the other hand,
for local algorithms, each ergodicity class contains only an
exponentially small fraction of the walks.
\bibitem{foot2}
It should be noted that the algorithm of Ref. \cite{Jagodzinski-etal_92}
is not ergodic.
\bibitem{foot3}
Note that
``possible" directions means here a direction for which there are
no intersections
with $\omega[i-1,i+2]$. It is worth mentioning that an \textsf{I}
(respectively \textsf{S}, \textsf{L}) always has $(2d-2)$
(respectively $(2d-4)$, $(2d-3)$) possible directions.
\bibitem{foot4}
This implementation of the reptation algorithm
has much in common with the hybrid algorithms used in simulations of
lattice field theories, see footnote 42 in Ref. \cite{Sokal_review}.
\end{thebibliography}
\end{document}